\documentclass[aoas,MSNbibl,nameyear,seceqn,dvips]{arximspdf}
\usepackage{dcolumn}
\usepackage{graphicx}

\doi{10.1214/11-AOAS473}
\volume{5}
\issue{3}
\pubyear{2011}
\firstpage{2024}
\lastpage{2051}

\makeatletter
\def\bptnote#1{}

\newcolumntype{d}[1]{D{.}{.}{#1}}
\newproclaim{remark}{Remark}
\newcommand{\cal}{\mathcal}

\renewcommand{\epsilon}{\varepsilon}

\newcommand{\bnu}{\bolds\nu}
\newcommand{\bu}{\mathbf{u}}
\newcommand{\bv}{\mathbf{v}}
\newtheorem{proposition}{Proposition}
\makeatother

\begin{document}
\begin{frontmatter}

\title{A space--time varying coefficient model: The equity of service
accessibility\thanksref{T1}}
\runtitle{A space--time varying coefficient model}

\begin{aug}
\author[A]{\fnms{Nicoleta} \snm{Serban}\corref{}\ead[label=e1]{nserban@isye.gatech.edu}}
\runauthor{N. Serban}
\affiliation{Georgia Institute of Technology}
\address[A]{H. Milton School of \\
\quad Industrial and Systems Engineering\\
Georgia Institute of Technology\\
765 Ferst Drive, NW\\
Atlanta, Georgia 30332-0205\\
USA\\
\printead{e1}} 
\end{aug}
\thankstext{T1}{Supported in part by the NSF Grant CMMI-0954283 and by
the industry research Georgia Tech foundation gift provided by Predictix.}

\received{\smonth{6} \syear{2010}}
\revised{\smonth{3} \syear{2011}}

%
\begin{abstract}
Research in examining the equity of service accessibility has emer\-ged
as economic and social equity advocates recognized that where people
live influences their opportunities for economic development, access to
quality health care and political participation. In this research paper
service accessibility equity is concerned with where and when services
have been and are accessed by different groups of people, identified by
location or underlying socioeconomic variables. Using new statistical
methods for modeling spatial-temporal data, this paper estimates
demographic association patterns to financial service accessibility
varying over a large geographic area (Georgia) and over a period of 13
years. The underlying model is a space--time varying coefficient model
including both separable space and time varying coefficients and
space--time interaction terms. The model is extended to a multilevel
response where the varying coefficients account for both the within- and
between-variability. We introduce an inference procedure for
assessing the shape of the varying regression coefficients using
confidence bands.
\end{abstract}

%

\end{frontmatter}

\section{Introduction}\label{sec;into}

\begin{quote}

``A home is more than a shelter---when located in a community with
resources and amenities it is a critical determinant of opportunity.''
[\citet{BlackwellandFox}]

\end{quote}

\textit{Service accessibility equity} is the study of systematic
disparities in a population's access to services that are considered
fundamental in fostering economic development, improving wellness and
enhancing the general quality of life of a population within a given
geographic area. Examples of such services are health care, education,
healthy food, financial services and others. \textit{Accessibility}
is\vadjust{\eject}
measured as utilization-scaled travel cost of a community $U$ to the
nearby sites in a service network consisting of multiple service sites
geographically distributed: ${\cal S} = \{s_1,\ldots,s_n\}$. A common
utilization measure is the population rate within the community and its
surroundings [\citet{Marsh94}]. In this paper the
utilization is measured as the population rate divided by the service
rate to account for the service availability for each population unit.

One challenge in measuring service accessibility is defining the travel
cost for the residents in a community to access the sites in the
service network. In the research works so far, the travel cost is
calculated as the average or minimum distance between the centroid of
the region $U$ and the nearby sites in the service network [\citet{Lovett};
\citet{Talen}; \citet{accessschool}]. However, communities occupy
uneven geographic areas varying in size, and, therefore, their
simplified representation by their centroids is restrictive. In this
research paper, we instead represent a community by a sample of
locations in the neighborhood $U$, $u_1,\ldots,u_B\in U$, and compute
the street-network distances from these sample locations to the service
network. Furthermore, the travel cost at each sample location $u_b$ is
measured as a summary of the travel distances, $\{d(u_b,s_i)\}
_{i=1,\ldots,n}$.

Combining the two ideas discussed above, utilization-adjustment of the
travel cost and representation of a community by a series of sample
spatial points, we evaluate the accessibility of a neighborhood to a
service network in year $t$ using
%
\begin{equation}\label{ac1}
Y(U,t) = \frac{1}{B}\sum_{b=1}^B (C(u_b,t)^{\beta} W(u_b,t) ),
\end{equation}
where $C(u_b,t)$ is the travel cost at the sample location $u_b$
measured as the average street-network distance to the closest $Q$
service sites available at time $t$ (in our study, $Q=3$), $W(u_b,t)$
is the utilization adjustment factor at location $u_b$ and $\beta$ is a
\textit{distance utility parameter}. We estimate $\beta$ by robust linear
regression: $\log(W(u_b,t))\sim-\log(C(u_b,t))$.

Dividing the geographic space into contiguous spatial units $U_{\!s},
s\,{=}\,1,\ldots,S$, where each spatial unit corresponds to a neighborhood
(e.g., census tract), the accessibility measure (utilization-adjusted
travel cost) varies across the geographic space and time; $Y(U_s,t) =
Y(s,t)$ defines the space--time varying accessibility process.
Moreover, there are multiple providers in the service network, the
accessibility process has an intrinsic multilevel structure. Under this
multilevel structure, let $Y_{p}(s,t)$ be the accessibility of the
community~$U_s$ to the sites of the $p$th provider for $p=1,\ldots,P$,
where $P$ is the number of service providers.

This research paper focuses on measuring and estimating
spatial-temporal patterns in the association between demographic
variables (including race, ethnicity and income) and service
accessibility. Specific questions that will be addressed within this\vadjust{\eject}
study are as follows: Is service accessibility equitable across
population groups varying in ethnicity and income? Do service
distribution inequities vary across regions and time? Are there service
providers that provide a more equitable distribution of their services
than others? What is the most common demographic feature associated
with inequities?

  To evaluate the equity of service accessibility with respect to
various population groups over a period of time and within a large
geographic space, we propose to estimate the space--time varying
association of the accessibility measure jointly over a series of
demographic variables. In this context, weak associations or the
absence of systematic disparities in service access are interpreted as
service accessibility equity. One challenge of this association
analysis is simultaneous estimation of the association patterns since
the goal is to assess both the equity with respect to various ethnicity
and race demographic variables controlling for income and the equity
with respect to income controlling for ethnicity and race. A second
challenge is that services are delivered within
a multilevel network - multiple providers which deliver across multiple service sites.

Many existing studies have analyzed service accessibility for different
groups of people identified by underlying socioeconomic variables, but
they are limited to small geographic areas and to only one year of data
[\citet{banks}; \citet{larson}; \citet{neighfoodstore2};
\citet{princeton}; \citet{accessschool}; \citet{assessacess};
\citet{zenk}].
Commonly employed statistical procedures include regression methods
assuming independence between service sites. Exploratory studies rely
on graphical diagnostics but not on statistical inference, which can be
used to make informed decisions. Although the methods applied to the
existing studies have usefulness for some research questions, a
spatial-temporal multivariate analysis of data with a multilevel
structure requires new statistical methods which are  rigorous, take
into account the dependence in the data, and implementable, apply to
real data complexity.

To this end, we introduce a space--time (multilevel) model which allows
estimation of space--time varying association patterns of a set of
functional predictors (e.g., demographic variables) to a functional
response, in our case study, the accessibility process. The modeling
procedure introduced in this paper falls under a more general
framework: varying-coefficient models. These models have been applied
to longitudinal data to estimate time-dependent effects on a response
variable [\citet{Assuncao}; \citet{varying3}; \citet{HastieandTibshirani};
\citet{varying1}; \citet{varying4}; \citet{WuandLiang};
\citet{varying7}]. \citet{waller07} review existing models to explore
space-varying regressions and propose a~Bayesian procedure.
\citet{Gelfandetal} briefly mention the extension of their proposed Bayesian
space-varying model to separable space--time varying coefficient models
with a warning on its computational challenges. Space--time
separability greatly simplifies the problem by reducing the
computational effort; however, it is a restrictive assumption since it
implies that dependence attenuates in a multiplicative/additive manner
across space and time. Therefore, extension of the Bayesian varying
coefficient model to more complex modeling (e.g., space--time
interaction) requires expensive computations which may be prohibitive
for densely sampled spatial domains.

Our methodological contribution is three-fold. First, we propose a
space--time varying coefficient model that takes into account the
interaction between time and space in a computationally efficient
manner. To overcome the computational complexity due to operations with
a large dependence matrix, we use a low-rank approximation to the
space--time coefficient processes using radial basis of functions
[\citet{semipar}]; this approach enables estimation of the
space--time varying coefficient model for densely observed space and/or
time domains.

Second, we extend this model to multilevel data, resulting in a
multilevel varying coefficient model. A few recent works have
considered the study of multilevel functional models
[\citet{Baladandayuthapanietal}; \citet{Crainiceanuetal};
\citet{Dietal}; \citet{MorrisandCarroll}; \citet{Morrisetal};
\citet{RiceandWu}; \citet{Staicu}; \citet{WuandZhang}]. In the related
research, models of multilevel functional data have been applied to
functional responses where the predictor is a fixed variate, commonly
time, and, more recently, they have been extended to functional
predictors but scalar responses. In this paper, the multilevel
functional model applies to both functional response and functional
predictors and it extends to the more difficult setting when the
functionality is with respect to space and time. Challenges in
estimating such a complex model include nonidentifiability and
computational efficiency. We overcome the identifiability problem by
using a knots-based kernel decomposition with a different set of knots
across the model coefficients. We use penalized splines for
computational efficiency in adapting to the smoothness in the
space--time varying coefficients [\citet{semipar}].

Third, we introduce an inference procedure to assess the shape of the
space--time varying coefficients. Generally, a common approach for
identifying the shape of a regression function is hypothesis testing.
However, for our model, hypothesis testing will require multiple tests
for deciding whether its shape is nonlinear, linear or constant as a
function of space or/and time. In this paper, we discuss an inference
procedure for assessing the shape of the varying regression
coefficients using confidence bands.

The rest of the paper is organized as follows. In Sections \ref{secmodel} and \ref{secmodelhierarchical} we present the space--time
varying coefficient model as well as its extension to multilevel data
along with the estimation and inference procedures. In Section \ref{seccasestudy} we present the application of the models introduced in
this paper to evaluate the equity of financial service accessibility in
Georgia. We first describe the data resources followed by the
discussion of our results and findings. Section \ref{secfinal}
concludes the paper. Some technical details are deferred to the
supplemental material [\citet{suppl}], which also provides
complemental graphical descriptions of our analysis of the equity of
service accessibility.

\vspace*{-2pt}\section{Space--time varying coefficient model}\vspace*{-2pt}
\label{secmodel}

\subsection{The model}\label{secmodeldefinition}

In this section we introduce a space--time varying coefficient model
for estimating the relationships between the accessibility process and
a series of demographic variables varying in time and space. The
observed data are $(Y_{ij},\{X_{r,ij}, r=1,\ldots,R\})$, where $Y_{ij}
= Y(t_i,s_j)$ is the response variable and $X_{r,ij} = X_r(t_i,s_j)$ a
set of covariates observed at location $s_j=(s_{j1},s_{j2})$,
$j=1,\ldots,S$, and time $t_i,  i=1,\ldots,T$, such that $\mathbb
{E}[Y_{ij}|X] =\gamma_{1}(t_i,s_j)X_{1,ij}+\cdots + \gamma
_{R}(t_i,s_j)X_{R,ij}$ where $\gamma_{r}(t,s)$ for $r=1,\ldots,R$ are
smooth coefficient functions. Note that not all covariates need to vary
in both time and space; the modeling procedure allows for various
predictor forms (scalar, varying in time, varying in space or both).
For example, in our model implementation we take $X_{1,ij} = 1$ and,
therefore, $\gamma_1(s,t)$ is the intercept coefficient.

In this paper we decompose the regression coefficients into separable
space and time global effects along with space--time deviations from
the global effects which are intrinsically local and account for the
interaction between space and time:
\begin{eqnarray*}
\gamma_r(t,s) = \alpha_r(t)+\beta_r(s)+\sum_{m=1}^{M_r}\sum
_{n=1}^{N_r}\nu_{r,mn}K_{\mathrm{temp}}\bigl(\bigl|t-\kappa^{(T)}_m\bigr|\bigr)K_{sp}\bigl(\bigl\|s-\kappa
^{(S)}_n\bigr\|\bigr).
\end{eqnarray*}
We decompose the global coefficient functions using the radial spline
basis [\citet{semipar}],
%
\begin{eqnarray}\label{eqglobaleffects}
\alpha_{r}(t) &=&
\tau_{r,0}+\tau_{r,1}t+\sum_{m=1}^{M_r} u_{r,m}K_{\mathrm{temp}}\bigl(\bigl|t-\kappa
^{(T)}_m\bigr|\bigr),\\
\beta_{r}(s_1,s_2) &=&
\delta_{r,0}+\delta_{r,11}s_1+\delta_{r,12}s_2+\sum_{n=1}^{N_r}
v_{r,n}K_{sp}\bigl(\bigl\|s-\kappa^{(S)}_n\bigr\|\bigr).
\end{eqnarray}
In these decompositions $K_{\mathrm{temp}}(t)$ is a temporal kernel whereas
$\kappa^{(T)}_m$, $m=1,\ldots,M_r$, are knots covering the time domain,
and $K_{sp}(s)$ is a spatial kernel whereas $\kappa^{(S)}_n$,
$n=1,\ldots,N_r$, are knots covering the space domain.

Importantly, although the kernel of the space--time interaction
coefficient is separable in time and space, the decomposition is not.
One advantage of using this kernel decomposition is that it allows
decomposition of the design matrix as a Kronecker product, which, in
turn, will ease the computations in the estimation procedure. We derive
the Kronecker product decomposition in
the
Supplemental Material 1 of this paper.\vadjust{\eject}

In the semiparametric literature a common kernel function is the radial
spline kernel function defined for $d$-dimensional domains [\citet{space-filling}].
Bivariate smoothing based on radial basis functions
has the advantage of being rotational invariant, which is important in
geographical smoothing. For two-dimensional domains, the function
$K_{sp}(\cdot)$ could be replaced by any other covariance function
[\citet{StatModel1993}], for example, the Mat\'{e}rn covariance function.

Knots for one-dimensional spaces are commonly set to the sample
quantiles of the observation points, whereas knots for two-dimensional
spaces are commonly selected using the space-filling algorithm [\citet{space-filling}],
which is based on minimax design, or $k$-nearest
neighbor clustering algorithms. In this paper we implement these
standard methods to select the number of knots.

\subsection{Estimation}\label{subsecestimation}

We choose a method for estimating the model described in the previous
section from among several candidate procedures. One modeling approach
is smoothing splines [\citet{Wahba}], which assumes that the number of
knots is equal to the number of observation design points ($M_r = T$
and $N_r=S$) and controls the smoothness of the coefficient by
penalizing the influence of the coefficients $u_{r,m}, m=1,\ldots,M_r$,
and $v_{r,n},n=1,\ldots,N_r$, using a penalty function. One primary
drawback of this estimation procedure is its computational aspect. A~less computational approach is regression splines [\citet{Wahba}], in
which a small number of knots are used ($M_r \ll T$ and $N_r\ll S$). This
reduces to selection of the optimal numbers of knots, which can be
computationally expensive in the context of our model since it requires
solving a multidimensional optimization problem. The smoothness levels
of the regression functions differ from one covariate to another and,
therefore, we need to optimally identify $(M_r, N_r)s$ for $r=1,\ldots
,R$. In addition, this approach introduces modeling bias.

An alternative approach to optimal knots selection is to assume equal
number of spatial knots ($N_r = N$ for $r=1,\ldots,R$) and equal number
of temporal knots ($M_r =M$ for $r=1,\ldots,R$) with $N$ and $M$
sufficiently large such that the modeling bias is small [\citet{Asymptotics}],
but, similarly to smoothing splines, impose constraints on the
coefficients $u_{r,m}$, $v_{r,n}$, and~$\nu_{r,nm}$ as follows:
\[
\sum_{m=1}^M u_{r,m}^2\leq C_{r}^{(T)}, \qquad   \sum_{n=1}^N v_{r,n}^2\leq
C_{r}^{(S)}, \qquad   \sum_{m=1}^M \sum_{n=1}^N \nu_{r,mn}^2\leq C_r
\]
or, equivalently, estimate the coefficients using penalized regression
\begin{eqnarray*}
&&\| h(Y_{k,ij}) - \gamma_{1}(t_i,s_j)X_{1,ij} -\cdots -\gamma
_{r}(t_i,s_j) X_{r, ij}\|^2\\
&&  \qquad {}+ \sum_{r=1}^R  \bigl\{\lambda_{r}^{(T)} \bu
_{r} \bu_{r}'+ \lambda_{r}^{(S)} \bv_{r} \bv_{r}'+\lambda_{r} \bnu_{r}
\bnu_{r}' \bigr\},
\end{eqnarray*}
where
\begin{eqnarray*}
\bu_{r} &=& (u_{r,1},\ldots,u_{r,N}), \qquad  \bv_{r} = (v_{r,1},\ldots
,v_{r,M}) \quad  \mbox{and }\\[2pt]
 \bnu_r &=& \{\nu_{r,nm}\}_{n=1,\ldots,N,m=1,\ldots,M}.
 \end{eqnarray*}
Moreover, the parameters $\lambda_{r}^{(T)}$, $\lambda_{r}^{(S)}$ and
$\lambda_{r}$ are penalties controlling the smoothness level of the
regression coefficients. This approach is often referred to as
penalized splines [\citet{semipar}]. Consequently, selection of
the number of knots reduces to selection of the penalty parameters,
which, in turn, is a multidimensional optimization problem.

In the semiparametric regression literature the problem of selecting
the penalties, and implicitly of the the number of knots, is overcome
by solving an equivalent mixed effects regression problem where $\bu
_{r}$, $\bv_{r}$ and $\bnu_r$ are random effects, specifically, $\bu
_{r}\sim N(0,(\sigma_{r}^{(T)})^2I_n)$, $\bv_{r}\sim N(0,(\sigma
_{r}^{(S)})^2I_m)$ and $\bnu_{r}\sim N(0,\sigma_{r}^2I_{nm})$. Under
the mixed effects model, the penalties are
\[
\lambda_{r}^{(T)} = \frac{\sigma_{\epsilon}^2}{(\sigma_{r}^{(T)})^2},
\qquad
\lambda_{r}^{(S)} = \frac{\sigma_{\epsilon}^2}{(\sigma_{r}^{(S)})^2}
 \quad \mbox{and}  \quad  \lambda_{r} = \frac{\sigma_{\epsilon}^2}{\sigma_{r}^2}.
\]
We therefore estimate the model parameters using a mixed effects model
to circumvent the difficulty of selecting the penalty parameters, or,
implicitly, the number of knots.

Based on the mixed-effects model formulation, denote the vector of the
fixed effects
\[
\bolds\Theta= \left[\matrix{\tau_{1,0} & \tau_{1,1} & \delta_{1,0} & \delta
_{1,11} & \delta_{1,12} &\cdots &\tau_{R,0} & \tau_{R,1} & \delta_{R,0}& \delta_{R,11} & \delta_{R,12}} \right]
\]
with identifiability constraints $\delta_{r,0}=0$ for $r=1,\ldots,R$.
The vector of random effects is
\[
\mathbf{U}= \left[\matrix{\bu_1 & \bv_1 & \bnu_1 &\cdots & \bu_R & \bv_R & \bnu
_R } \right].
\]
The corresponding design matrices ${\cal X}$ and ${\cal Z}$ are
\begin{eqnarray}
{\cal X} &=&  \left[\matrix{\mathbf{X}_1(t_i,s_j) &\cdots & \mathbf
{X}_R(t_i,s_j)} \right]_{i=1,\ldots,T,j=1,\ldots,S} \nonumber \\[2pt]
 \eqntext{ \mbox{with }
 \mathbf
{X}_r(t_i,s_j)=X_r(t_i,s_j) \left[\matrix{1 & t_i & 1 & s_{1j} & s_{2j}}\right ],} \\[2pt]
{\cal Z} &=&  \left[\matrix{\mathbf{Z}_1(t_i,s_j) &\cdots & \mathbf
{Z}_R(t_i,s_j)} \right]_{i=1,\ldots,T,j=1,\ldots,S}\nonumber \\[2pt]
 \eqntext{ \mbox{with }
\mathbf{Z}_r(t_i,s_j)=X_r(t_i,s_j)   \left[\matrix{K_{\mathrm{temp}}\bigl(\bigl|t_i-\kappa
^{(T)}_{m}\bigr|\bigr)&  K_{sp}\bigl(\bigl\|s_i-\kappa^{(S)}_{n}\bigr\|\bigr)\hphantom{ii}}\right.}\\[2pt]
\eqntext{ \left.\matrix{
K_{\mathrm{temp}}\bigl(\bigl|t_i-\kappa^{(T)}_{m}\bigr|\bigr)& K_{sp}\bigl(\bigl\|s_i-\kappa^{(S)}_{n}\bigr\|\bigr)}\right ].}
\end{eqnarray}
%
The model in the matrix form becomes $\mathbb{E}[Y|{\cal X}]={\cal
X}\bolds\Theta+{\cal Z}\mathbf{U}$, which is equivalent to a
linear mixed model.

\subsection{Inference}\label{secinference}

In this section we discuss alternative methods for making inference on
the shape of the regression coefficients entering the space--time
varying coefficient model described in Section \ref{secmodeldefinition}. Specifically, we discuss a procedure for
evaluating the shape (constant vs. linear vs. nonlinear) of the
temporal and spatial global coefficients and a procedure for testing
the significance of the space--time interaction.

\textit{Shape evaluation.} In this section we discuss a novel procedure
for shape evaluation of the temporal regression coefficients. A similar
procedure applies to spatial regression coefficients. Specifically,
each temporal regression coefficient can take various shapes, for
example, constant [$\alpha(t) = \tau_{0}$], linear [$\alpha(t) = \tau
_{0}+\tau_{1}t$] or nonlinear.

In varying-coefficient models, the common procedure for assessing the
shape of the coefficients is hypothesis testing. For example, the
hypothesis test for linearity is equivalent to $H_0\dvtx  \sigma^2_u= 0
\mbox{ vs. }  H_a\dvtx  \sigma^2_u > 0$ where $\sigma_u^2$ is the variance
of the random effects $u_{m}$ under the mixed effects model. The common
approach is a likelihood ratio testing (LRT) procedure. \citet{Crainiceanuetal2005}
developed a LRT by taking advantage of the existing research
in hypothesis testing for zero variance in linear mixed-effects (LME)
models. \citet{Liangetal} tested for linearity of nonparametric
functions using a Cr\'{a}mer--von Mises statistic.

Although there are several competitive approaches for testing for
linearity of the regression coefficients, because we need to test
sequential hypotheses to decide about the shape of a coefficient and
because we often have a large number of predictors that enter the
space--time varying coefficient model, we instead propose identifying
the shape of the coefficients using simultaneous confidence bands. If
$CB_{\gamma}$ is a $1-\gamma$ confidence band for the coefficient~$\alpha(t)$,
then $P(\alpha(t)\in CB_{\gamma},  t\in{\cal T})\geq
1-\gamma$ where ${\cal T}$ is the time domain. The derivation of the
joint confidence bands is presented in the
Supplemental Material~2 of this paper.

Many authors have noted that using confidence intervals has a series of
advantages over the conventional hypothesis testing [\citet{SimandReid}].
Confidence intervals cannot only be used to test a
hypothesis, but also to provide additional information on the
variability of an observed sample statistic and on its probable
relationship to the value of this statistic in the population from
which the sample was drawn.

%
\begin{figure}[t]

\includegraphics{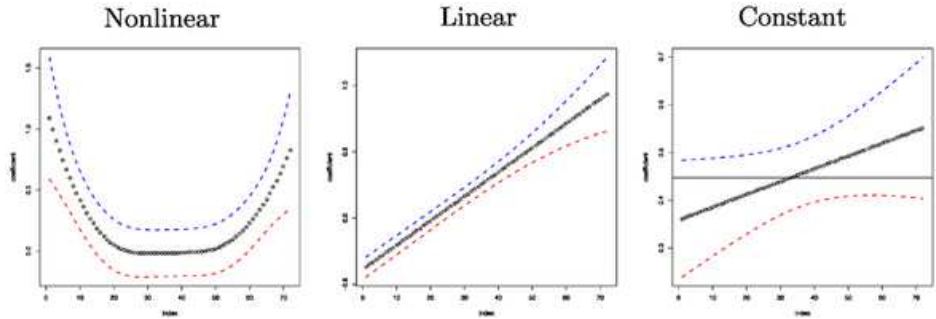}

\caption{Examples of nonlinear, linear, constant effects.} \label{3examples}
\end{figure}

Figure \ref{3examples} depicts examples of three different
one-dimensional shapes along with their confidence bands (CB). We
define ``constant'' shape if there exists a constant line that falls
within the confidence bands. Similarly, we define ``linear'' shape if
there exists a linear function that falls within the confidence bands.
When searching for a line $L$ within the confidence bands, it suffices
to search for linear functions between the convex hull of the upper
level of the confidence band and the convex hull of the lower level of
the confidence band.   Although the coverage of the shape test
hypothesis is maintained when using confidence bands, the power may be
reduced, as we point out in the simulation study included in
Supplemental Material~4.

According to the result below, accepting the null hypothesis at a
significance level $\gamma$ is equivalent to finding at least one null
shape function in the set of all possible functions in the $(1-\gamma)$
confidence band.

\begin{proposition} Denote $\Theta_c = \{\alpha(t)\dvtx   \alpha
(t)=c,  c\in\mathbb{R}\}$ (the set of all real constant functions).
The rejection rule of the hypothesis test for constant shape ($H_0\dvtx
\alpha(t)\in\Theta_c$) becomes
\begin{eqnarray*}
\Theta_c\cap CB_{\gamma} = \varnothing \qquad \mbox{where }   P\bigl(\alpha(t)\in
CB_{\gamma}\bigr) = 1-\gamma.
\end{eqnarray*}
Using this rejection rule, the type I error is equal to $\gamma$.
\end{proposition}

This proposition follows from the classical result on the equivalence
of confidence intervals and hypothesis testing in \citet{Lehmann}.

\textit{Space--time interaction.} In our modeling approach, in order to
account for the space--time interaction, we introduce an additional
term in $\gamma_r(t,s)$ specified by a set of effects $\bnu_r = \{\nu
_{r,nm}\}_{n=1,\ldots,N,m=1,\ldots,M}$. The testing procedure for
space--time interaction of the regression coefficient for the $r$th
predictor reduces to
\[
H_0\dvtx   \sigma_{\nu_r} = 0  \quad \mbox{vs.}  \quad  H_0\dvtx    \sigma_{\nu_r} > 0.
\]
The null hypothesis implies that the association between the $r$th
predictor and the response is separable in time and space.

Although there are several approaches for testing the null hypothesis
of  zero variance component in linear mixed effects models,
hypothesis tests that apply under multiple variance components have
been investigated only recently. To test for space--time interaction
under a multiple predictor model, we therefore use the approximations
to the finite sample null distribution of the RLRT statistic in \citet{Grevenetal}.

\section{Multilevel varying coefficient model}\label{secmodelhierarchical}

In this section we discuss the extension of the varying-coefficient
model in Section \ref{secmodel} to data with an intrinsic multilevel structure.

\vspace*{3pt}\subsection{The model}\label{secmodeldefinitionhierarchical}

The observed data for the $p$th category (e.g., service provider) for
$p=1,\ldots,P$ are $(Y_{p, ij},X_{r,ij})$, where $Y_{p, ij} =
Y_p(t_i,s_j)$ is a generalized response variable and $X_{r,ij} =
X_r(t_i,s_j)$ the $r$th covariate observed at location
$s_j=(s_{j1},s_{j2})$ and time $t_i$ with $\mathbb{E}[Y_{p,ij}|X] =
\gamma_{1p}(t_i,s_j)X_{1,ij}+\cdots +\gamma_{Rp}(t_i,s_j)X_{R,ij}$ where
$\gamma_{rp}(t,s)$ for $r=1,\ldots,R$ are smooth coefficient functions.
In our application, $\{Y_{p, ij}\}_{i=1,\ldots,T, j=1,\ldots,S}$ are
the measures of service accessibility to the $p$th service provider sites.

To assess the association deviations of each of the $k$th group of
processes from the global association pattern, we further decompose the
regression coefficients as follows:
%
\begin{eqnarray}\label{e6}
\gamma_{rp}(t_i,s_j) =\gamma_{r}(t_i,s_j)+\eta_{rp}(t_i,s_j),
\end{eqnarray}
where $\gamma_{r}(t,s)$ specifies the global association patterns and
$\eta_{rp}(t_i,s_j)$ specifies the group-specific deviations from the
global association patterns. We further assume that the global effects
$\gamma_{r}(t,s)$ take an additive form
\[
\gamma_{r}(t,s) =\alpha_{r}(t)+\beta_{r}(s),
\]
where the time- and space-varying regression coefficients follow the
decomposition in (\ref{eqglobaleffects}). We also assume that the
group-specific regression coefficients are decomposed according to
\begin{eqnarray*}
\eta_{rp}(t,s) = \alpha_{rp}(t)+\beta_{rp}(s) + \sum_{m=1}^{M}\sum
_{n=1}^{N}\nu_{r,p,nm}K_{\mathrm{temp}}\bigl(t-\kappa^{(T)}_{p,m}\bigr)K_{sp}\bigl(s-\kappa
^{(S)}_{p,n}\bigr),
\end{eqnarray*}
where $\alpha_{rp}(t)$ and $\beta_{rp}(s)$ are decomposed using the
radial spline basis similarly to the formulas in (\ref{eqglobaleffects}). We denote $\kappa^{(T)}_{p,m}$, $m=1,\ldots,M$,
the temporal knots used in the decomposition of the time-varying
regression coefficient and $\kappa^{(S)}_{p,n}$, $n=1,\ldots,N$, the
spatial knots used in the decomposition of the space-varying regression
coefficient for the $p$th service provider. For example, the
decomposition of the regression coefficient $\alpha_{rp}(t)$ is
\[
\alpha_{rp}(t) =
\tau_{rp,0}+\tau_{rp,1}t+\sum_{m=1}^{M} u_{rp,m}K_{\mathrm{temp}}\bigl(\bigl|t-\kappa
^{(T)}_{p,m}\bigr|\bigr).
\]

\subsection{Estimation}\label{subsecestimhierarchical}

Similar to the varying coefficient model in Section \ref{secmodel}, we
estimate the parameters in the multilevel varying coefficient model
using the mixed effects model equivalence, resulting in a multilevel
mixed effects model.

For the multilevel model, we need to impose a series of constraints on
the fixed effects and on the selection of the temporal and spatial
knots. For $r=1,\ldots,R$,
\begin{eqnarray*}
\sum_{p=1}^P \tau_{rp,0} &=& 0 \quad  \mbox{and}  \quad  \sum_{p=1}^P \tau_{rp,1} =
0, \\
\sum_{p=1}^P \delta_{rp,0} &=& 0,  \qquad   \sum_{p=1}^P \delta_{rp,1} =
0\quad
\mbox{and}  \quad  \sum_{p=1}^P \delta_{rp,12} = 0.
\end{eqnarray*}
\begin{proposition}\label{prop2}  If the temporal and spatial knots are
selected such that
%
\begin{eqnarray*}
\bigl|\kappa_{m_1,p}^{(T)}-\kappa_{m_2,p'}^{(T)}\bigr|> d^{(T)}
\end{eqnarray*}
for any $m_1,m_2 \in\{1,\ldots,M\}$, and for any $p, p'=0,1,\ldots,P\ (p\neq p')$,
\begin{eqnarray*}
\bigl\|\kappa_{n_1,p}^{(S)}-\kappa_{n_2,p'}^{(S)}\bigr\|> d^{(S)}
\end{eqnarray*}
for any $n_1,n_2 \in\{1,\ldots,N\}$, and for any $p, p'=0,1,\ldots,P\ ( p\neq p')$,
where $d^{(T)}$ and $d^{(S)}$ are away from zero, then the model
parameters in the multilevel model decomposition in Section \ref{secmodeldefinitionhierarchical} are identifiable.
\end{proposition}

The proof of this proposition is provided in Supplemental Material 3 of
this paper.

\subsection{Inference}

Since making inference under the multilevel model presented in the
previous section implies making inference over all groups jointly, we
need to correct for multiplicity. For instance, given that we need to
evaluate the shape of the temporal global effects in the decomposition
of the time-varying coefficients corresponding to the $r$th predictor,
$\alpha_{rp}(t)$ for $p=1,\ldots,P$, we test multiple hypotheses [e.g.,
$H_{0p}\dvtx   \alpha_{rp}(t)   \mbox{ constant}$] simultaneously. For a
small number of groups ($P$ small), we can modify the approach
discussed in Section \ref{secinference} to account for the joint
inference. Consequently, we estimate joint confidence bands:
\[
P(\alpha_{rp} \in CB_{rp},  p=1,\ldots,P) \geq1-\rho
\]
by correcting the confidence level of individual confidence bands for
multiple inference using a Bonferroni correction; that is, estimate
$1-\rho/P$ confidence bands. Under the classical definition of the type
I error for joint inference, we find that the test using $1-\rho$ joint
confidence bands is $\rho$,
\[
\mbox{type I error} = \sum_{p=1}^P \operatorname{Pr}_{H_{rp}}(\Theta_{\mathrm{const}}\cap
CB_{rp}=\varnothing)\leq\rho.
\]
Note that this correction will provide overly conservative confidence
band estimates when $P$ is large.

\section{Case study: The equity of financial services}\label{seccasestudy}

We proceed with the application of the varying coefficient models to
assess whether there are systematic disparities in the service
accessibility with respect to various demographic variables. We focus
on the equity of financial service accessibility in the state of
Georgia over a period of 13 years, 1996--2008. We start with a
description of the accessibility data followed by a brief exploratory
analysis of the demographic variables. We continue with the
presentation of the findings from the application of the varying
coefficient models.

\subsection{Accessibility data}\label{secdata}

The site location data in this study were acquired from the Federal
Deposit Insurance Corporation (FDIC). In our study we use data starting
from 1996 to 2008. We geocoded the site location addresses using ArcGIS
(ESRI) to obtain the service point locations in the service network:
${\cal S} = \{s_1,\ldots,s_n\}$ ($n=2\mbox{,}849$ for Georgia).

In service research the distance between a service site and its
customers is commonly evaluated using the Euclidean or the Manhattan
distance between the centroid of the neighborhood and the location of
the closest service site. GIS road network data allows including more
realistic route distances. For example, Talen (\citeyear{Talen,accessschool}) uses the
street-network distance to compute the distance between the centroid of
the neighborhood and the site location. Lovett et al. (\citeyear{Lovett}) use road
distance and travel time by car.  We acquired highway data for the
whole U.S. (courtesy of the GIS Center at Georgia Institute of
Technology) as well as a TIGER street-detailed network for Georgia and
we took the average of the travel distances computed using both
networks to obtain the distances $d(u_b,s_i; t)$ for $b=1,\ldots,B$
sample locations within a community and $s_1,\ldots,s_n$ service sites.
Notably, none of the two networks provide highly accurate travel
distances; therefore, the average over the distances computed using the
two networks will provide more robust distance estimates. Finally, the
travel cost $C(u_b,t)$ is computed as the average over the smallest
three distances in $\{d(u_b,s_i; t), i=1,\ldots,n\}$.

Last, we obtain the utilization adjustment weights using the population
counts acquired from the Environmental Systems Research Institute
(ESRI). We use kernel smoothing [Diggle (\citeyear{D1985})] to estimate the rate of
point spatial processes. Using this approach, we obtain the population
and service rate estimates at the sample locations, $P(u_b, t),
b=1,\ldots,B$ and $R(u_b,t), b=1,\ldots,B$, in year $t$. Further, we
compute the utilization weights using $W(u_b,t) = P(u_b, t)/R(u_b,t),
b=1,\ldots,B$, and along with the travel cost $C(u_b,t)$, we can
finally obtain the accessibility measure at the community level using
equation (\ref{ac1}). We apply this estimation procedure for all
communities in Georgia and obtain the accessibility process $Y(U_s,t)$
for $s=1,\ldots,S$ ($S=1\mbox{,}624$) and $t=1996,\ldots,2008$.   In this
research, census tracts are used as proxy for communities. According to
the Census Bureau, census tracts are delineated with local input and
intended to represent neighborhoods.

  \begin{remark*}  Since the accessibility measure is an adjusted travel
cost, we interpret it as follows. Large values of the travel cost or
large values of the measure correspond to low accessibility to the
service network. Therefore, if the measure values are, for example,
increasing, the access to service is decreasing. Moreover, if the
association of a demographic variable to the proposed measure is high,
we infer that there is low accessibility with respect to the
demographic variable.
\end{remark*}

\begin{figure}

\includegraphics{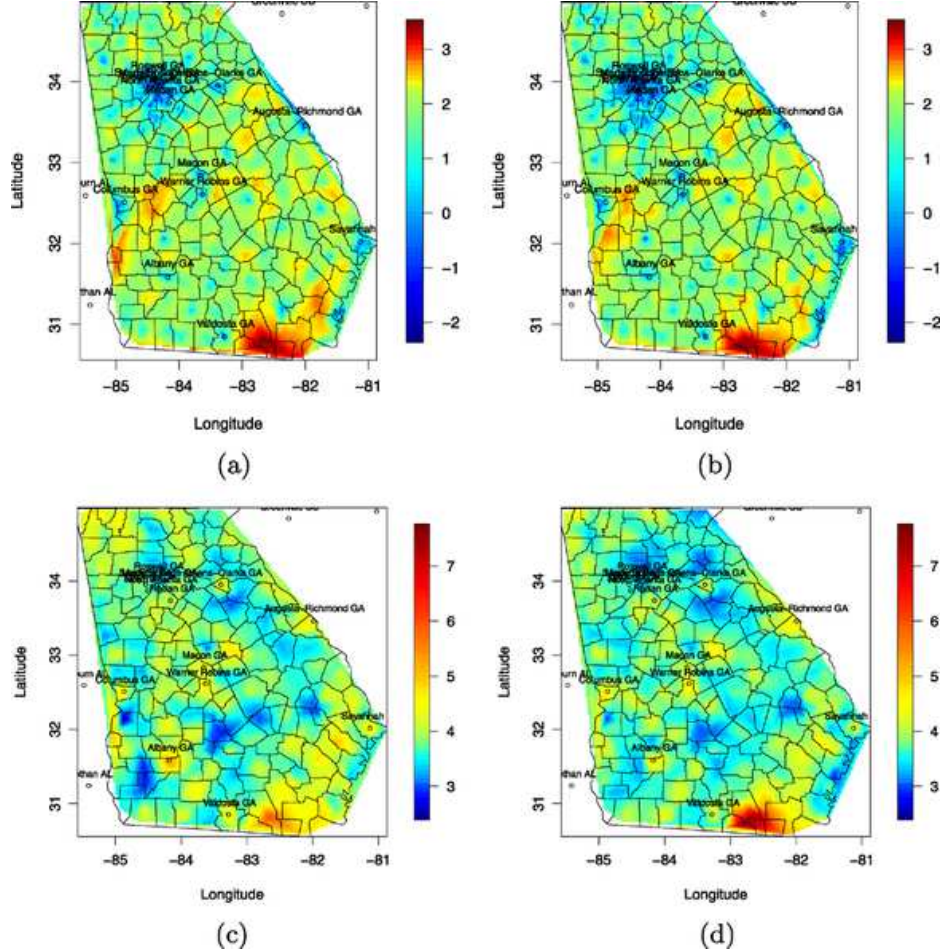}

\caption{\textup{(a)} Logtravel cost---1996; \textup{(b)} Log travel cost---2008;
\textup{(c)} Log accessibility---1996; and \textup{(d)}~Log accessibility---2008.}
\label{figtravelcost}
\end{figure}

In the following discussion, we contrast ``\textit{horizontal equity}''
[Figure \ref{figtravelcost}(a), (b)] measured using the travel cost
without adjusting for the ``utilization'' of a service operation [in
equation (\ref{ac1}), $W(u,t)=1$ for any location $u$ and time point
$t$] to ``\textit{vertical equity}'' [Figure \ref{figtravelcost}(c), (d)]
which accounts for the expected utilization of a service.

Although difficult to assess visually, there are more extensive areas
with lower (unadjusted) travel cost in 2008 than in 1996 in Georgia. On
the other hand, the access to financial services is slightly lower in
2008 than in 1996 for highly populated regions, more specifically,
Atlanta (see
Supplemental Material 6 for the travel cost maps of metropolitan
Atlanta). The primary reason for this contrast is that the increase in
the number of new financial sites has a lower slope than the population
growth in highly populated regions in Georgia. Consequently, these
regions have weaker access compared to low density population areas,
although the travel cost is small. These findings point to potential
business opportunities for financial service providers.

This comparison between travel cost without and with
utilization-adjust\-ment motivates the need for correcting the travel
cost for the expected utilization of a service. In our subsequent
analysis, we will only focus on the utilization-adjusted travel cost.

\subsection{Demographics data}

In this study the demographic variables used to predict service
accessibility include median household income, race and ethnicity data
which are acquired from the Environmental Systems Research Institute
(ESRI).   The description of the methodology employed to obtain the
demographic estimates at the census tract level is provided in
Supplemental Material 5 of this paper. One has to bear in mind that the
demographic estimates are measured with error which, in turn, will
impact the estimates of the association between accessibility and the
demographic variables.

\begin{figure}

\includegraphics{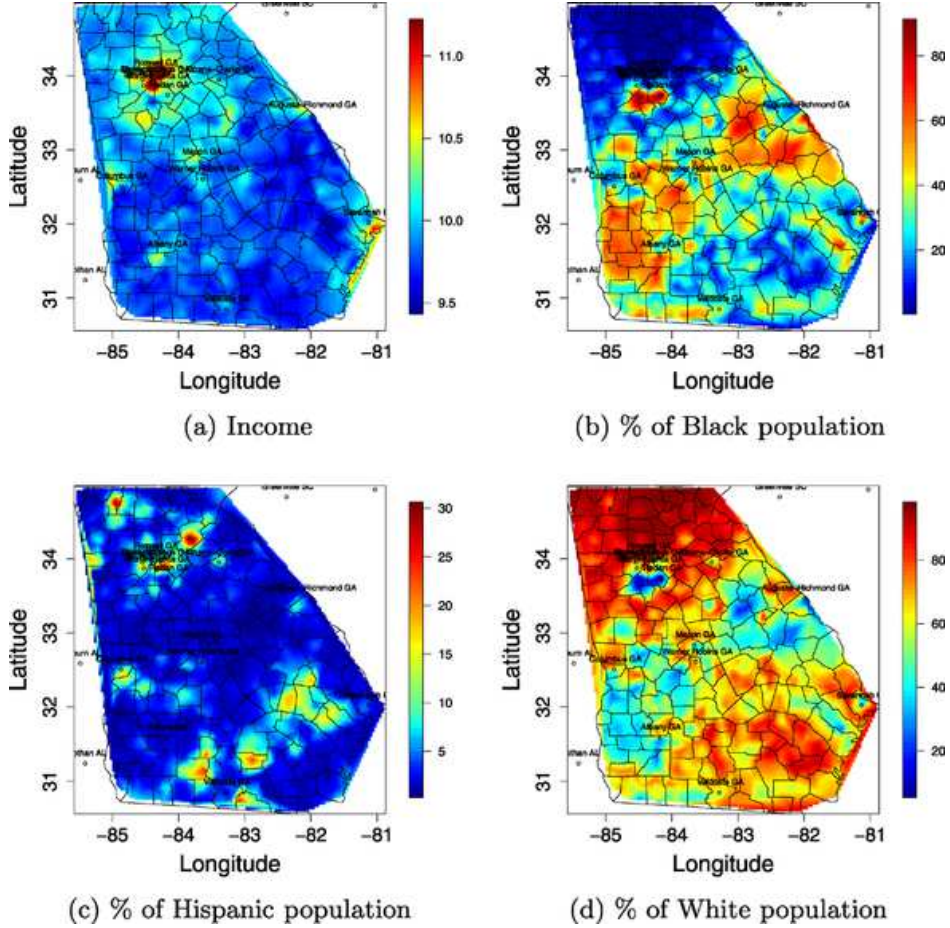}
\vspace*{-2pt}
\caption{Demographic variables in Georgia.}
\label{figdemo}
\vspace*{-6pt}
\end{figure}

Since the boundaries of census tracts are updated by the Census Bureau
every ten years, our data set includes a change of boundaries. The
Census Bureau provides the so-called ``relationship files'' to document
the revisions of the 1990 to 2000 census tract boundaries. We map the
data collected before 1999 to 2000 boundaries using the information in
these relationship files.

Figure \ref{figdemo} shows the income level on the log scale and the
percentages of Black, Hispanic and White populations for the last year
of demographic data in this study. We do not show the plot of Asian
percentages since overall in Georgia (except a small area in Atlanta)
the percentage of Asian population is very low (close to zero).
Contrasting the plots displaying the percentage of Black and White
populations, we note that areas of high Black population have low White
population and vice versa, pointing to significant segregation between
black and white populations in Georgia. Indeed, the correlation is as
high as $-0.98$, which suggests high collinearity between these two
variables. At the same time, the collinearity between any other two
demographic variables is low (see Table \ref{tabt1}). Since there is
high collinearity only between White and Black populations, we fit the
varying coefficient models separately for $ \{\mbox{income,
Hispanic, White} \}$ and for $ \{\mbox{income, Hispanic,
Black} \}$.


\begin{table}
\tabcolsep=0pt
\caption{Correlation between demographic variables}
\vspace*{-2pt}
\label{tabt1}
\begin{tabular*}{160pt}{@{\extracolsep{\fill}}ld{2.2}d{2.2}d{2.2}@{}}
\hline
\textbf{Income} & \multicolumn{1}{c}{\textbf{White}} & \multicolumn{1}{c}{\textbf{Black}}
& \multicolumn{1}{c@{}}{\textbf{Hispanic}} \\
\hline
\hphantom{$-$}1.00 & 0.18 & -0.27 & 0.37 \\
\hphantom{$-$}0.18 & 1.00 & -0.98 & -0.01 \\
$-$0.27 & -0.98 & 1.00 & -0.16 \\
\hphantom{$-$}0.37 & -0.01 & -0.16 & 1.00 \\
\hline
\end{tabular*}
\end{table}

\subsection{Varying coefficient model: Motivation, results and
findings}\label{secVCMresults}

In this paper we introduce a framework for studying the equity of
service accessibility for different groups of people identified by
location or underlying socioeconomic variables. The data consist of a
series of maps characterizing the access to financial services and a
series of maps describing the demographic composition at the
neighborhood level varying in time. The objective is to assess
geographically-varying association patterns between accessibility and
demographic variables over a period of several years.

Simple visual inspection of a large number of maps (13 for the
accessibility measure and $13\times4$ for four demographic variables)
observed over a large geographic space goes beyond feasibility.
Moreover, the existing models will only allow partial understanding of
the dynamics in the equity of service accessibility. For example,
space-varying coefficient models provide a one-year snapshot of the
equity in service accessibility but will neither explain how it has
changed over time nor account for the interaction between space and
time associations. To model the space--time dynamics in the association
between accessibility and socioeconomic variables jointly, we therefore
apply the space--time varying coefficient model in Section \ref{secmodel}.

In the application of the space--time varying coefficient model to the
data in this study, we selected a small number of temporal basis
functions \mbox{($M=7$)} since we have a small number of time points; the
space--time varying coefficients do not change significantly for
various values of $M$. However, the estimated space--time varying
coefficients vary with the number of spatial basis functions, $N$. For
small~$N$, the space-varying coefficients are smooth. \citet{Ruppert}
empirically suggests that after a minimum number of knots has been
reached, the modeling bias is small. Therefore, we can control the
modeling bias by using a large enough~$N$; in our application $N$ can
be as large as $S = 1\mbox{,}624$. In contrast, the larger $N$ is, the more
expensive the computation is. Consequently, we need to select $N$ for
an optimal trade-off between modeling bias and computational
feasibility. To select~$N$, we used a residual-based analysis suggested
by \citet{wood}.

When interpreting the varying regression coefficients, one has to bear
in mind that large values of the accessibility measure
(population-adjusted travel cost) correspond to weak access to
financial services. Moreover, significant association between
accessibility and a demographic variable suggests that access to
financial services is driven in part by the presence or the absence of
the population group identified by the corresponding
variable.\looseness=1

\begin{figure}[t!]

\includegraphics[scale=0.99]{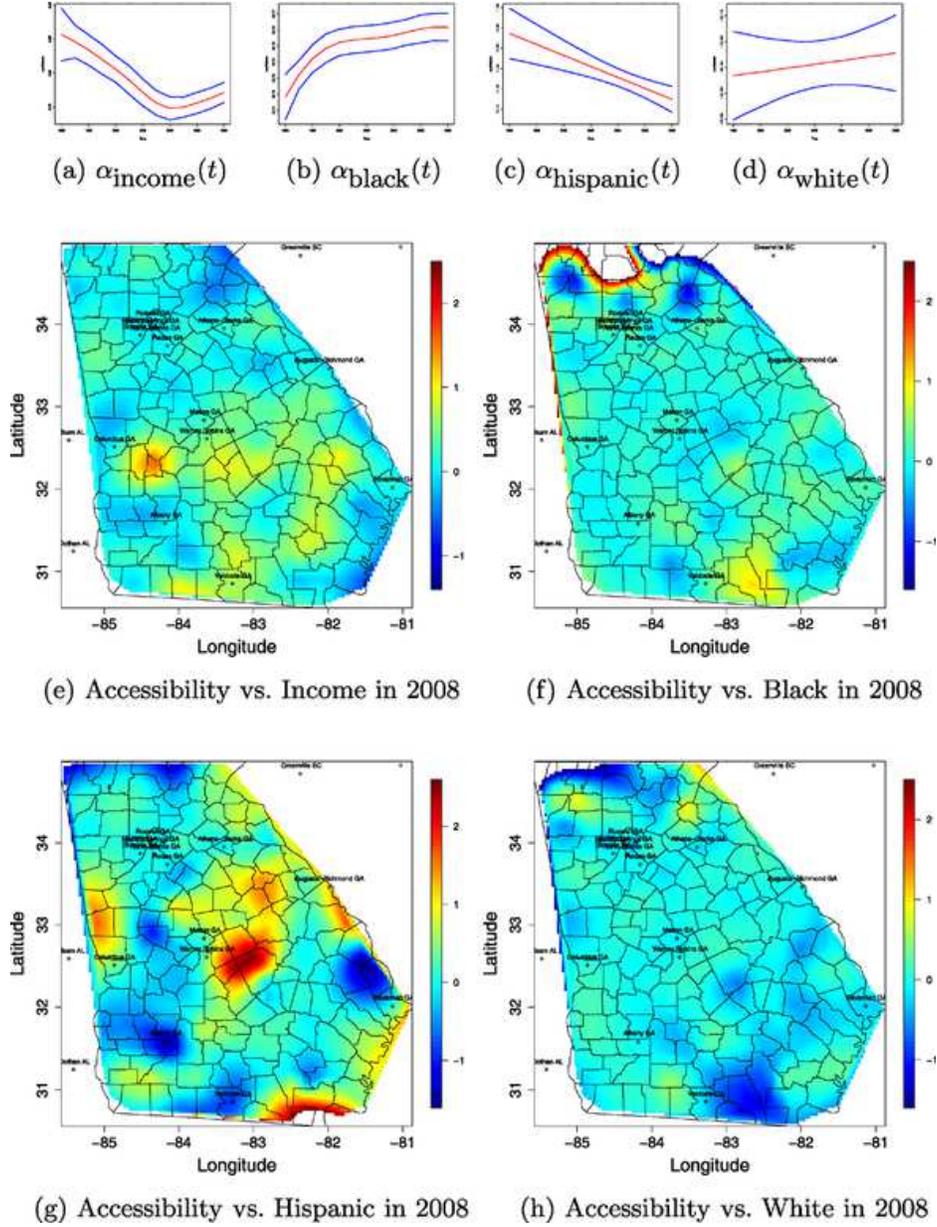}%
\vspace*{-3pt}
\caption{\textup{Georgia}: The time-varying coefficients \textup{(a)}--\textup{(d)} and the
spatial relationship pattern in 2008 for four demographic covariates---income,
percentage of Black, Hispanic and White populations.}
\label{figVCMcoefficientsgeorgia}
\vspace*{-6pt}
\end{figure}

\begin{figure}[t!]

\includegraphics{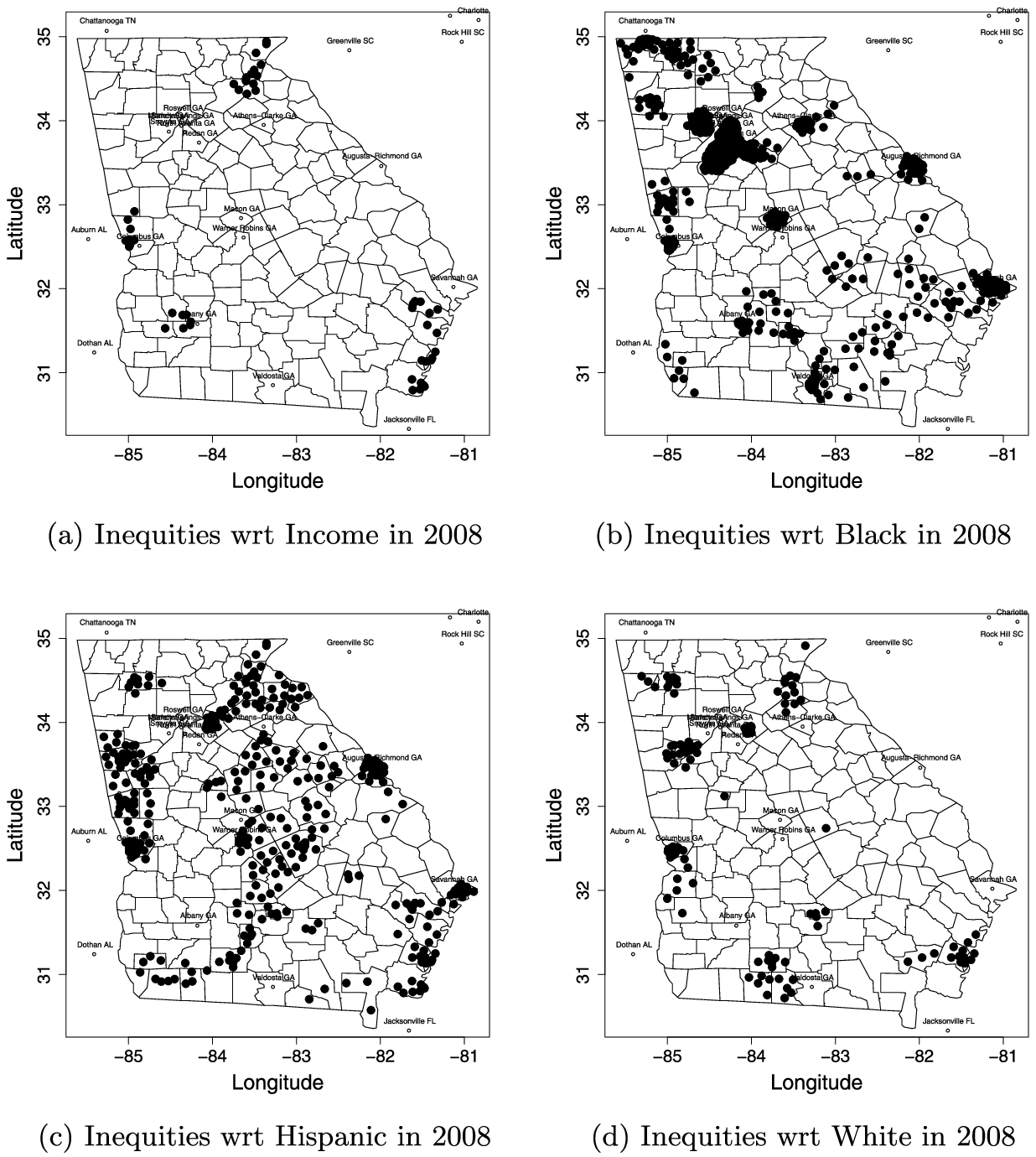}

\caption{\textup{Georgia}: Inequity locations in 2008 with respect to
(wrt) four demographic covariates---income, percentage of Black,
Hispanic and White populations.}
\label{figVCMineqsgeorgia}
\end{figure}

\begin{figure}[t!]

\includegraphics{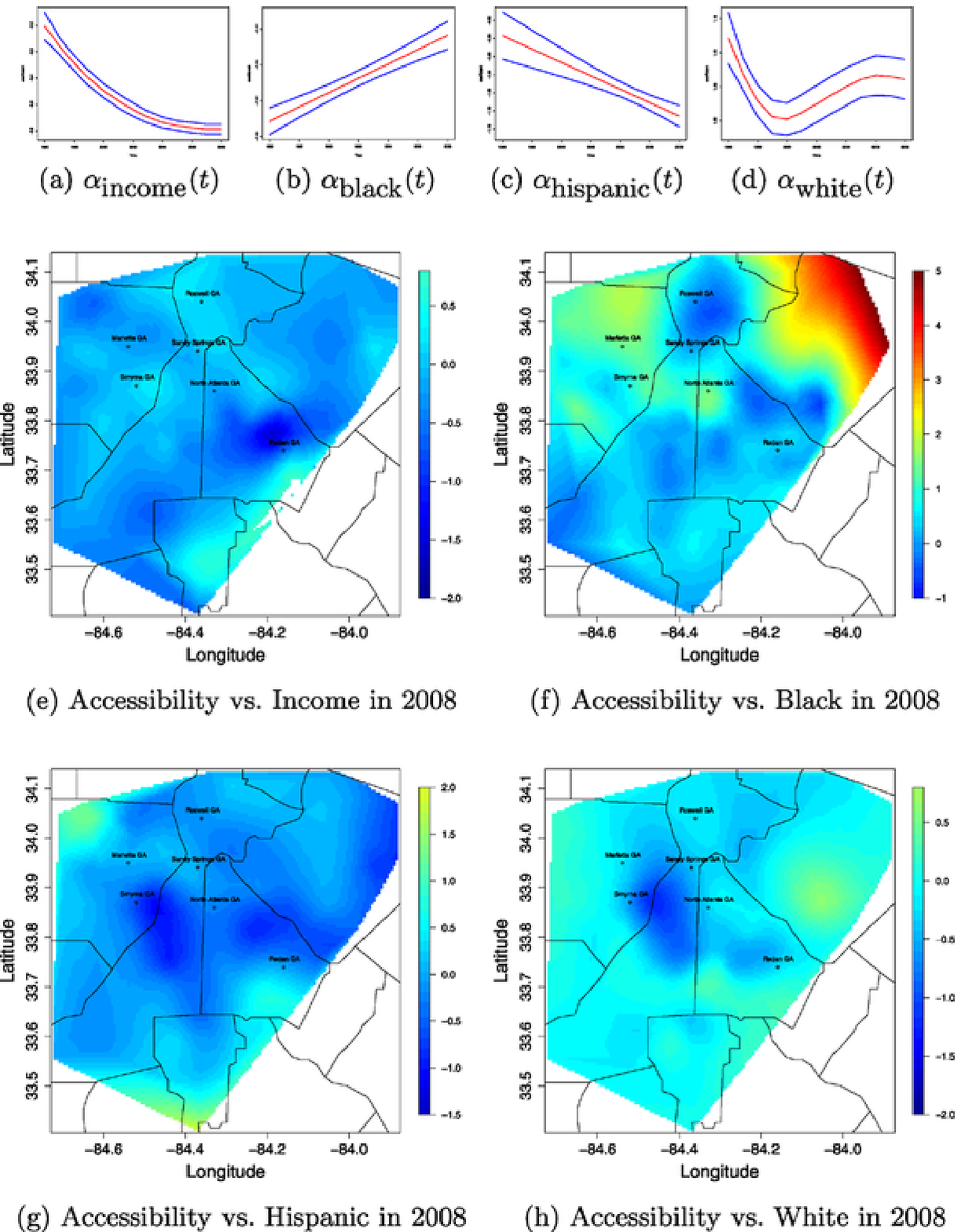}
\vspace*{-4pt}
\caption{\textup{Atlanta}: The time-varying coefficients \textup{(a)}--\textup{(d)} and the
spatial relationship pattern in 2008 for four demographic covariates---income,
percentage of Black, Hispanic and White populations.}\vspace*{-1pt}
\label{figVCMcoefficientsatlanta}
\vspace*{-6pt}
\end{figure}

\begin{figure}

\includegraphics{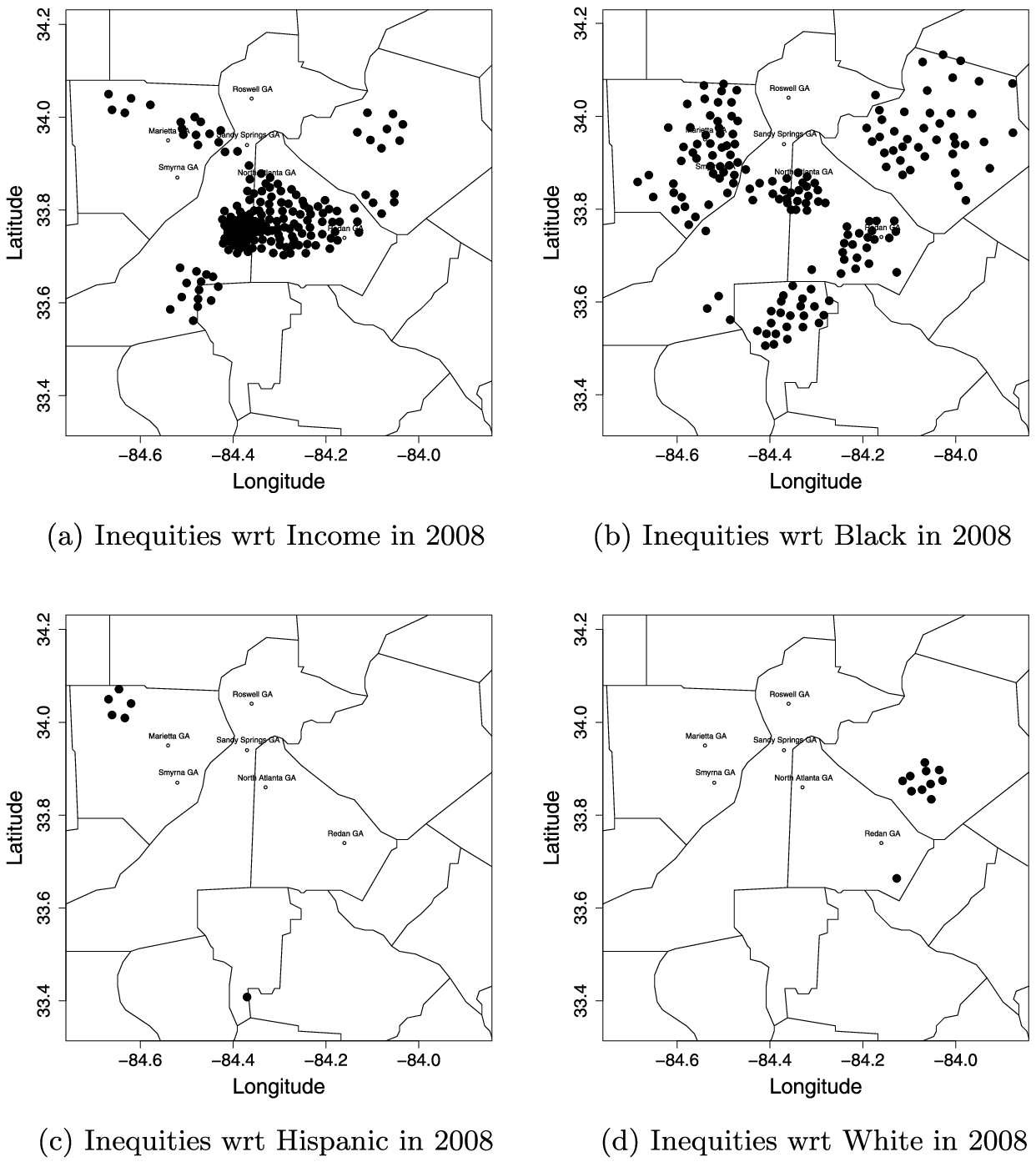}

\caption{\textup{Atlanta}: Inequity locations in 2008 with respect to (wrt)
four demographic covariates---income, percentage of Black, Hispanic and
White populations.}
\label{figVCMineqsatlanta}
\end{figure}

In this section we summarize our findings based on Figures \ref{figVCMcoefficientsgeorgia},
\ref{figVCMineqsgeorgia}, \ref{figVCMcoefficientsatlanta} and \ref{figVCMineqsatlanta}, which
include the following:
\begin{longlist}[(3)]
\item[(1)] The time-varying coefficients for income, \% of Black population,
\% of Hispanic population and \% of White population.

\item[(2)] The space-varying association patterns for the four
demographic\break \mbox{covariates} in 2008 calculated from $\gamma(2008,s) = \alpha(2008)+\beta
(s)+\break\operatorname{Interaction}(2008,s)$.

\item[(3)] The point locations of inequities with respect to the four
demographic covariates in 2008.
\end{longlist}

The output figures summarize the space--time relationships between
accessibility and the socioeconomic variables considered in this study.
We highlight\vadjust{\eject} that without a rigorous modeling procedure, we cannot
evaluate the significance of the associations to service accessibility.
Therefore, using the space--time varying model is important not only
for estimation of these associations but also for inference about their
significance as described below.

We define locations of inequity with respect to income to be the
spatial units $s$ such that $\gamma_{\mathrm{income}}(2008,s)$ is
statistically significantly positive (positive correlation between
income and utilization-adjusted travel cost). We also define locations
of inequity with respect to race/ethnicity (percentage of Black,
Hispanic and White populations) to be the spatial units $s$ such
that~$\gamma_{\mathrm{ethnicity}}(2008,s)$ is statistically significantly
negative (negative correlation between percentage and
utilization-adjusted travel cost). Statistical significance of the
coefficients is derived from the simultaneous confidence bands of the
spatial coefficients. Specifically, the coefficient at\vadjust{\eject} location $s$
is statistically significantly positive with 95\% significance level if
the lower bound of the confidence interval at $s$ is positive and it is
statistically significantly negative if the upper bound of the
confidence interval at $s$ is negative.

The time-varying coefficients corresponding to each demographic
covariate and their confidence bands are in Figure \ref{figVCMcoefficientsgeorgia}(a)--(d) and Figure
\ref{figVCMcoefficientsatlanta}(a)--(d). Using the approach for
evaluating the shape of the varying coefficients in Section \ref{secinference}, we infer that at the significance level of $95\%$, the
time-varying coefficients for income are nonlinear, for the \% of Black
and Hispanic populations are linear whereas for the \% of White
population are constant. All space-varying coefficients are nonlinear.

\begin{table}
\tabcolsep=0pt
\caption{$P$-values for testing the significance of the space--time
interaction in the varying coefficients for four demographic variables.}
\label{table1}
\begin{tabular*}{\textwidth}{@{\extracolsep{\fill}}lcccc@{}}\hline
\textbf{Covariate} & \textbf{Income} & \textbf{\% of Black} & \textbf{\% of Hispanic} & \textbf{\% of White} \\
\hline
$p$-value & $0.0018$ & $0.0009$ & $\approx$0 & $\approx$0 \\\hline
\end{tabular*}
\end{table}

We also evaluate the significance of the interaction terms for all four
demographic variables. We apply the testing procedure for the
space--time interaction term described in Section \ref{secinference}.
The $p$-values are provided in Table \ref{table1}.\vadjust{\goodbreak} For all four
demographic variables, the space--time interaction terms are highly
significant, which implies that the space--time interactions in the
association patterns are statistically significant; this suggests that
the interaction term has a significant contribution to the spatial
association patterns. Therefore, using a varying coefficient model with
a space--time interaction term significantly contributes to a more
accurate association analysis.

\textit{Time-varying association patterns.} Following the inference
procedure for shape evaluation, we infer that over the past 13 years in
the state of Georgia, the association between the access to financial
services and two demographic variables, income and the percentage of
Hispanic population, has strengthened over time with a brief decrease
in the last years, whereas the association between the access to
financial services and the percentage of Black population has weakened.
This suggests that access to financial services has become more and
more dependent on the income level of the residents in a~community and
whether they are of Hispanic descent but less dependent on the race of
the population.

Importantly, we cannot make inference about the magnitude of the
association patterns since the constants for the temporal and spatial
coefficients,~$\tau_{0}$ and $\delta_{0}$, are nonidentifiable. For
inference on the level and the direction (negative or positive) of the
service accessibility association we need to investigate the
space-varying association year by year as discussed below.

\textit{Space-varying association patterns.} Controlling for race and
ethnicity, the association between access to financial services and
income level varies throughout the state of Georgia, with primarily
weak positive association in the north but negative association in the
south   [Figure \ref{figVCMcoefficientsgeorgia}(e)]. This pattern
is consistent with the income map in Figure \ref{figdemo}(a); the
income is consistently low in south and middle Georgia except for a few
urban areas. This suggests that regions with low income population tend
to also have lower access to financial services regardless of race and
ethnicity.  Moreover, there are only a few locations with
statistically significant positive association between income and
utilization-adjusted travel cost [Figure \ref{figVCMineqsgeorgia}(a)].
This implies that although south Georgia consists primarily of
low income population whereas north Georgia is more mixed with higher
income population than south, financial services are present in both.

 The map of the Black population percentage is not as uniform as its
association to accessibility; there is a high density of the Black
population in south Atlanta and in mid to south Georgia but not in the
north [Figure \ref{figdemo}(b)].  On the other hand, there are
several locations with statistically significant inequities as shown in
Figure \ref{figVCMineqsgeorgia}(b), although the association of
the \% of Black population to financial service access is weak in
Georgia except for the upper north. Most of these locations are in
urban areas. We therefore conclude that the inequities in access to
financial services with respect to the Black population are present but
low throughout Georgia.

 The association between travel cost and the \% of Hispanic population
is neither uniformly positive nor high [Figure \ref{figVCMcoefficientsgeorgia}(g)]. Areas of high Hispanic density
population have low but statistically significantly positive
association [Figure \ref{figVCMineqsgeorgia}(c)]. This indicates
that the presence of financial services decreases with the increase in
Hispanic population.

In contrast, the association for White population is consistently weak
throughout Georgia, although the White population density is high in
most of Georgia except in the middle [Figure \ref{figdemo}(d)].
Moreover, there are much fewer inequity locations than for Black and
Hispanic populations and most are in rural areas [Figure~\ref{figVCMineqsgeorgia}(b)--(d)].

Since Atlanta is the largest city in Georgia with mixed income
population and with a high percentage of Black, Hispanic and White
populations, we applied the modeling procedures proposed in this paper
to evaluate potential inequities in the Atlanta area and its surroundings.

The only time-varying coefficient that changes its shape in comparison
to Georgia is for the White population; it has a nonlinear shape. There
is an increase in the impact of the \% of White population on the
access to services (equivalently, a decrease in the impact on the
travel cost) up to 2000 followed by a slower decrease thereafter.

Significant inequities in the Atlanta area are with respect to income
and the Black population. The association of the \% of Black population
to service access is negative and strong in many communities in south
and north Atlanta, implying significant inequities even after
controlling for the income level. {However, the association map does
not fully overlap with the density of the Black population; that is,
while South Atlanta has a large Black population [Figure 4(b),
Supplemental Material 6], most inequities are in North Atlanta.

There is a positive association between income and access to financial
services in south Atlanta, an area with a predominantly low income
population [Figure 4(a),
Supplemental Material 6]. Moreover, there is a weak association in
north Atlanta and negative association in the east and west borders
(possibly over-served areas). Therefore, when comparing the association
map and its statistical significance to the map of the per capita
income [Figure 4(a),
Supplemental Material 6], we conclude that many communities with low
and median income in central Atlanta have low access to financial
services, as there is a statistically significant association between
utilization-adjusted travel cost and income in these communities.

The association pattern for the \% of Hispanic and \% of White
population is largely negative, with just a few communities with
statistically significant positive association, although both
population groups are well represented in Atlanta [Figure 4(c), (d),
Supplemental Material 6], indicating insignificant inequities with
respect to the Hispanic and White populations in Atlanta.

\subsection{Multilevel varying coefficient model: Motivation, results
and findings}\label{secHVCMresults}

Since monopoly is not common in service distribution, there is an
intrinsic multilevel structure to service accessibility. At a higher
level, we estimate the association to service accessibility over all
service providers, whereas at a lower level, we estimate the deviations
from the overall patterns associated with each service provider in the
network. One simple approach would be to apply the space--time varying
coefficient model to the accessibility measure computed for each
service provider separately. However, this approach only takes into
account the variability within the network of each service provider but
not the variability between service providers; in other words, this
simple approach does not allow estimating the deviations from the
between-providers association patterns. The multilevel varying
coefficient model in Section \ref{secmodelhierarchical} estimates
both overall space--time association patterns and the deviations from
the overall pattern corresponding to each service provider.\looseness=1

In this section we discuss the association accessibility patterns to
five financial service providers: Bank of America (BoA), Branch Banking
and Trust Company (BB\&T), Regions Bank, SunTrust Bank and Wachovia
(Wells Fargo since 2008). In Supplemental Material 7 of this paper, we
include the corresponding association patterns derived from the
application of the multilevel space--time varying coefficient models.

All five banks are in the top 10 largest banks in the U.S., with a
variety of financial services including retail and commercial banking,
mortgages, insurance products, trust services and securities brokerage.
SunTrust and Regions banks are mainly based in southern states, BB\&T
is a national bank, whereas BoA and Wachovia are national banks with
international subsidiaries. Although Bank of America has dominated the
financial service market for many years, due to mergers and
acquisitions, SunTrust Bank and Wachovia (Wells Fargo since 2008) Bank
now dominate the market. The only bank that has not increased the
number of brunches in Georgia and, in fact, has closed some of them, is
Bank of America.

\textit{Time-varying association patterns.}   Following the inference
procedure for shape evaluation, we infer that the time-varying
deviations from overall association patterns denoted in this paper by
$\eta_{rp}(t_i,s_j)$, where $r$ is the index for the service provider
and $p$ is the predictor index, are all approximately zero (not
statistically significant) except for the deviation coefficients
corresponding to income. Therefore, over the past 13 years in the state
of Georgia, the association between the access to financial services
and income is positive and has strengthened for BB\&T and Bank of
America beyond the global upward trend, indicating stronger association
between service access and the income level in 2008 as compared to
1996. The income deviation coefficient for Wachovia is decreasing over
time but negative, implying a decrease in association with respect to
income. Finally, there are not significant systematic disparities in
the provider-specific deviations from the overall association patterns
for the ethnicity and race demographic variables.

\textit{Space-varying association patterns.} The lowest association
between service access and the four demographic variables is for Bank
of America, Regions and Wachovia. The accessibility association
patterns for these three banks do not deviate significantly from the
global trends. The association of the percentage of Black and White
populations to service accessibility is approximately zero, whereas the
association to the percentage of Hispanic population is weak with mixed
association throughout Georgia.

In urban and rural Georgia, the association between the access to BB\&T
services and the income level is highly positive, whereas the
association for the \% of Black and Hispanic populations is
consistently weak and for the~\% of White population is approximately
zero. Notably, the most significant inequities in 2008 for BB\&T are
with respect to income and they have increased over the 13 year period.
For SunTrust, the service accessibility association to income is strong
in south Georgia. There is weaker association in urban areas than in
rural areas. Moreover, the association of the \% of Black and Hispanic
populations is weak.

The most significant inequities in Atlanta are with respect to income---high
association between income and service accessibility for Bank of
America and Regions in south Atlanta, for BB\&T and SunTrust throughout
Atlanta. There are contrasting association patterns for north and south
Atlanta which also differ in their demographic decomposition.
Generally, the accessibility association is low for the White
population after controlling for income; this suggests equitable
accessibility to financial services for the White population.

\section{Final considerations}\label{secfinal}

The methodological contributions described in this paper are twofold.
First, we introduce a framework for the study of the equity of service
accessibility across population groups with various demographic
characteristics. This study allows characterization of the
geographically varying equity patterns over a period of several years.
Second, we investigate spatio-temporal estimation methods, which use
the underlying structure of varying coefficient models. The first model
estimates space--time varying association to a response variable (e.g.,
accessibility measure) of a~series of predictors (e.g., demographic
variables) jointly. The second model extends the first model to a
response variable with a multilevel structure. Because of the
complexity of the model parameters, we propose a simplified inference
procedure based on confidence bands which allows evaluation of the
shape of the varying coefficients.

We note that different service accessibility measures will provide
different accessibility maps, and, therefore, different conclusions
will be drawn for the study of service accessibility equity. In this
paper, the underlying measure is defined as the utilization-adjusted
travel cost; in Section \ref{secdata} we compared the accessibility
maps with and without correction for utilization and we concluded that
the two measures will provide different perspectives in the equity of
service accessibility, primarily for areas with low density population.
A comparison study of accessibility measures is beyond the scope of
this paper.

From the analysis of service accessibility using the space--time
varying coefficient model, we identified significant but low inequities
in some regions of Georgia with respect to income after controlling for
race and ethnicity, and with respect to Black and Hispanic populations
after controlling for income. These inequities have increased over
time. The most predominant inequities in Atlanta are for Black
population, although they have decreased over time. The association
between income and service access is largely positive in Atlanta,
suggesting potential inequities with respect to income as well. After
accounting for service utilization, there are more significant
inequities in urban areas than in rural areas; this may be due to the
fact that the population in rural areas is more homogeneous.

In the analysis of service accessibility using the multilevel
space--time model, we found for Georgia, and particularly for Atlanta,
that the deviations from the between-provider association patterns are
very insightful. Specifically, we learn, for example, that
income-driven inequities for BB\&T are significantly stronger when
contrasted to the overall association to accessibility, whereas for
other service providers, for example, Bank of America and Wachovia
(Wells Fargo), there are not significant deviations from the
between-providers association patterns.

Importantly, one challenge in space--time varying coefficient model
estimation is whether an assumed pattern in a multiple predictor model
can actually be recovered. For this, we conducted a simulation study
with two predictors. The association patterns for both predictors are
nonseparable in space and time, a more realistic simulation framework.
For this simulation, the estimated coefficients are accurately
estimated. We also evaluated the coverage and the power of the shape
evaluation procedure discussed in Section \ref{secinference}. The
power is lower for the time-varying components than for the
space-varying ones; in this simulation study, the number of spatial
points is $S=300$, whereas the number of time points is $T=15$, which
may lead to lower accuracy in the shape evaluation of the varying
coefficients. Moreover, the power depends on how smooth and close to
the null hypothesis the shape function is.
}

We note that different service accessibility measures will provide
different accessibility maps, and, therefore, different conclusions
will be drawn for the study of service accessibility equity. In this
paper the underlying measure is defined as the utilization-adjusted
travel cost; in Section \ref{secdata} we compared the accessibility
maps with and without correction for utilization and we concluded that
the two measures will provide different perspectives in the equity of
service accessibility. A comparison study of various accessibility
measures is beyond the scope of this paper.

One limitation of the study of service accessibility equity for large
geographic regions (e.g., the US) using the space-varying coefficient
models introduced in this paper is the computational aspect. Although
we have reduced the estimation of multiple space--time varying
coefficients with different smoothing levels to a simple mixed effects
model, the estimation problem remains computationally intensive. A
large geographic space requires a~large number of knots, which in turn
results in a model with a large number of random effects. In addition,
because of the model decomposition into separable and nonseparable
space--time coefficients, the number of variance components is large
even for a small number of predictors (e.g., for three predictors in
our study, we have a total of 12 variance components for the simple
varying coefficient model but as many as 42 for the multilevel model).
Ongoing research focuses on overcoming these challenges by using a
backfitting estimation algorithm in the presence of multiple
predictors.\looseness=1

\section*{Acknowledgments}
The author is grateful to Sungil Kim for assisting in implementing the
methods introduced in this paper, to Huijing Jiang for preparing some
of the data used in this project and to Jessica Heier Stamm for
providing insightful suggestions on improving the presentation of this
paper. The author is thankful to the two referees who provided very
useful feedback, and to the Associate Editor whose input greatly helped
improving the presentation of this paper.

\begin{supplement}
\stitle{Supplemental Material}
\slink[doi]{10.1214/11-AOAS473SUPP} 
\slink[url]{http://lib.stat.cmu.edu/aoas/473/supplement.pdf}
\sdatatype{.pdf}
\sdescription{The supplemental materials accompanying this paper are divided into
seven sections:
\\\indent
\textbf{Supplement 1.} Varying-coefficient model---Decomposition of the
design matrix under the tensor-product decomposition of the space--time
varying coefficients.
\\\indent
\textbf{Supplement 2.}  Varying-coefficient model---Derivation of the
confidence bands for the space and time varying coefficients.
\\\indent
\textbf{Supplement 3.}  Varying-coefficient model---A simulation study
under multiple predictors.
\\\indent
\textbf{Supplement 4.}  Varying-coefficient model---Proof of Proposition~\ref{prop2}.
\\\indent
\textbf{Supplement 5.} Case study---Description of ESRI data.
\\\indent
\textbf{Supplement 6.} Case study---Accessibility maps for Atlanta
area.
\\\indent
\textbf{Supplement 7.} Case study---Results and maps for the
provider-level accessibility analysis.}
\end{supplement}


\printaddresses


\begin{thebibliography}{44}

\bibitem[\protect\citeauthoryear{Assuncao}{2003}]{Assuncao}
\begin{barticle}[auto:STB|2011-03-03|12:04:44]
\bauthor{\bsnm{Assuncao},~\bfnm{R.~M.}\binits{R.~M.}}
(\byear{2003}).
\btitle{Space varying coefficient models for small area data}.
\bjournal{Environmetrics}
\bvolume{14}
\bpages{453--473}.
\end{barticle}
\endbibitem

\bibitem[\protect\citeauthoryear{Baladandayuthapani et al.}{2008}]{Baladandayuthapanietal}
\begin{barticle}[mr]
\bauthor{\bsnm{Baladandayuthapani},~\bfnm{Veerabhadran}\binits{V.}},
  \bauthor{\bsnm{Mallick},~\bfnm{Bani~K.}\binits{B.~K.}},
  \bauthor{\bsnm{Hong},~\bfnm{Mee~Young}\binits{M.~Y.}},
  \bauthor{\bsnm{Lupton},~\bfnm{Joanne~R.}\binits{J.~R.}},
  \bauthor{\bsnm{Turner},~\bfnm{Nancy~D.}\binits{N.~D.}} \AND
  \bauthor{\bsnm{Carroll},~\bfnm{Raymond~J.}\binits{R.~J.}}
(\byear{2008}).
\btitle{Bayesian hierarchical spatially correlated functional data analysis
  with application to colon carcinogenesis}.
\bjournal{Biometrics}
\bvolume{64}
\bpages{64--73, 321--322}.
\bid{doi={10.1111/j.1541-0420.2007.00846.x}, issn={0006-341X}, mr={2422820}}
\end{barticle}
\endbibitem

\bibitem[\protect\citeauthoryear{Blackwell and Fox}{2004}]{BlackwellandFox}
\begin{bmisc}[auto:STB|2011-03-03|12:04:44]
\bauthor{\bsnm{Blackwell},~\bfnm{A.~G.}\binits{A.~G.}} \AND
  \bauthor{\bsnm{Fox},~\bfnm{R.~K.}\binits{R.~K.}}
(\byear{2004}).
\bhowpublished{Regional equity and smart growth: Opportunities for advancing
  social and economic justice in America. \textit{PolicyLink and Funders'
  Network for Smart Growth and Livable Communities}.
  PolicyLink,
  Oakland, CA}.
\end{bmisc}
\endbibitem

\bibitem[\protect\citeauthoryear{Crainiceanu, Staicu and Di}{2009}]{Crainiceanuetal}
\begin{barticle}[pbm]
\bauthor{\bsnm{Crainiceanu},~\bfnm{Ciprian~M.}\binits{C.~M.}},
  \bauthor{\bsnm{Staicu},~\bfnm{Ana-Maria}\binits{A.-M.}} \AND
  \bauthor{\bsnm{Di},~\bfnm{Chong-Zhi}\binits{C.-Z.}}
(\byear{2009}).
\btitle{Generalized multilevel functional regression}.
\bjournal{J. Amer. Statist. Assoc.}
\bvolume{104}
\bpages{1550--1561}.
\bid{doi={10.1198/jasa.2009.tm08564}, issn={0162-1459}, mid={NIHMS127980},
  pmcid={2897156}, pmid={20625442}}
\bptnote{check year}%
\end{barticle}
\endbibitem

\bibitem[\protect\citeauthoryear{Crainiceanu et al.}{2005}]{Crainiceanuetal2005}
\begin{barticle}[mr]
\bauthor{\bsnm{Crainiceanu},~\bfnm{Ciprian}\binits{C.}},
  \bauthor{\bsnm{Ruppert},~\bfnm{David}\binits{D.}},
  \bauthor{\bsnm{Claeskens},~\bfnm{Gerda}\binits{G.}} \AND
  \bauthor{\bsnm{Wand},~\bfnm{M.~P.}\binits{M.~P.}}
(\byear{2005}).
\btitle{Exact likelihood ratio tests for penalised splines}.
\bjournal{Biometrika}
\bvolume{92}
\bpages{91--103}.
\bid{doi={10.1093/biomet/92.1.91}, issn={0006-3444}, mr={2158612}}
\end{barticle}
\endbibitem

\bibitem[\protect\citeauthoryear{Cressie}{1993}]{StatModel1993}
\begin{bbook}[mr]
\bauthor{\bsnm{Cressie},~\bfnm{Noel A.~C.}\binits{N.~A.~C.}}
(\byear{1993}).
\btitle{Statistics for Spatial Data}.
\bpublisher{Wiley}, \baddress{New York}.
\bid{mr={1239641}}
\end{bbook}
\endbibitem

\bibitem[\protect\citeauthoryear{Di et al.}{2009}]{Dietal}
\begin{barticle}[mr]
\bauthor{\bsnm{Di},~\bfnm{Chong-Zhi}\binits{C.-Z.}},
  \bauthor{\bsnm{Crainiceanu},~\bfnm{Ciprian~M.}\binits{C.~M.}},
  \bauthor{\bsnm{Caffo},~\bfnm{Brian~S.}\binits{B.~S.}} \AND
  \bauthor{\bsnm{Punjabi},~\bfnm{Naresh~M.}\binits{N.~M.}}
(\byear{2009}).
\btitle{Multilevel functional principal component analysis}.
\bjournal{Ann. Appl. Statist.}
\bvolume{3}
\bpages{458--488}.
\bid{doi={10.1214/08-AOAS206}, issn={1932-6157}, mr={2668715}}
\end{barticle}
\endbibitem

\bibitem[\protect\citeauthoryear{Diggle}{1985}]{D1985}
\begin{barticle}[auto:STB|2011-03-03|12:04:44]
\bauthor{\bsnm{Diggle},~\bfnm{P.}\binits{P.}}
(\byear{1985}).
\btitle{A kernel method for smoothing point process data}.
\bjournal{J. R. Stat. Soc. Ser. C Appl. Stat.}
\bvolume{34}
\bpages{138--147}.
\end{barticle}
\endbibitem

\bibitem[\protect\citeauthoryear{Fan and Zhang}{2000}]{varying3}
\begin{barticle}[mr]
\bauthor{\bsnm{Fan},~\bfnm{Jianqing}\binits{J.}} \AND
  \bauthor{\bsnm{Zhang},~\bfnm{Jin-Ting}\binits{J.-T.}}
(\byear{2000}).
\btitle{Two-step estimation of functional linear models with applications to
  longitudinal data}.
\bjournal{J. R. Stat. Soc. Ser. B Stat. Methodol.}
\bvolume{62}
\bpages{303--322}.
\bid{doi={10.1111/1467-9868.00233}, issn={1369-7412}, mr={1749541}}
\end{barticle}
\endbibitem

\bibitem[\protect\citeauthoryear{Gelfand et al.}{2003}]{Gelfandetal}
\begin{barticle}[mr]
\bauthor{\bsnm{Gelfand},~\bfnm{Alan~E.}\binits{A.~E.}},
  \bauthor{\bsnm{Kim},~\bfnm{Hyon-Jung}\binits{H.-J.}},
  \bauthor{\bsnm{Sirmans},~\bfnm{C.~F.}\binits{C.~F.}} \AND
  \bauthor{\bsnm{Banerjee},~\bfnm{Sudipto}\binits{S.}}
(\byear{2003}).
\btitle{Spatial modeling with spatially varying coefficient processes}.
\bjournal{J. Amer. Statist. Assoc.}
\bvolume{98}
\bpages{387--396}.
\bid{doi={10.1198/016214503000170}, issn={0162-1459}, mr={1995715}}
\end{barticle}
\endbibitem

\bibitem[\protect\citeauthoryear{Graves}{2003}]{banks}
\begin{barticle}[auto:STB|2011-03-03|12:04:44]
\bauthor{\bsnm{Graves},~\bfnm{S. M.}\binits{S. M.}}
(\byear{2003}).
\btitle{Landscapes of predation, landscapes of neglect: A location analysis of
  payday lenders and banks}.
\bjournal{The Professional Geographer}
\bvolume{55}
\bpages{303--317}.
\end{barticle}
\endbibitem

\bibitem[\protect\citeauthoryear{Greven et al.}{2008}]{Grevenetal}
\begin{barticle}[mr]
\bauthor{\bsnm{Greven},~\bfnm{Sonja}\binits{S.}},
  \bauthor{\bsnm{Crainiceanu},~\bfnm{Ciprian~M.}\binits{C.~M.}},
  \bauthor{\bsnm{K{\"u}chenhoff},~\bfnm{Helmut}\binits{H.}} \AND
  \bauthor{\bsnm{Peters},~\bfnm{Annette}\binits{A.}}
(\byear{2008}).
\btitle{Restricted likelihood ratio testing for zero variance components in
  linear mixed models}.
\bjournal{J. Comput. Graph. Statist.}
\bvolume{17}
\bpages{870--891}.
\bid{doi={10.1198/106186008X386599}, issn={1061-8600}, mr={2649072}}
\end{barticle}
\endbibitem


\bibitem[\protect\citeauthoryear{Hastie and Tibshirani}{1993}]{HastieandTibshirani}
\begin{barticle}[mr]
\bauthor{\bsnm{Hastie},~\bfnm{Trevor}\binits{T.}} \AND
  \bauthor{\bsnm{Tibshirani},~\bfnm{Robert}\binits{R.}}
(\byear{1993}).
\btitle{Varying-coefficient models}.
\bjournal{J. Roy. Statist. Soc. Ser. B}
\bvolume{55}
\bpages{757--796}.
\bid{issn={0035-9246}, mr={1229881}}
\bptnote{check related}%
\end{barticle}
\endbibitem

\bibitem[\protect\citeauthoryear{Hoover et al.}{1998}]{varying1}
\begin{barticle}[mr]
\bauthor{\bsnm{Hoover},~\bfnm{Donald~R.}\binits{D.~R.}},
  \bauthor{\bsnm{Rice},~\bfnm{John~A.}\binits{J.~A.}},
  \bauthor{\bsnm{Wu},~\bfnm{Colin~O.}\binits{C.~O.}} \AND
  \bauthor{\bsnm{Yang},~\bfnm{Li-Ping}\binits{L.-P.}}
(\byear{1998}).
\btitle{Nonparametric smoothing estimates of time-varying coefficient models
  with longitudinal data}.
\bjournal{Biometrika}
\bvolume{85}
\bpages{809--822}.
\bid{doi={10.1093/biomet/85.4.809}, issn={0006-3444}, mr={1666699}}
\end{barticle}
\endbibitem

\bibitem[\protect\citeauthoryear{Huang, Wu and Zhou}{2002}]{varying4}
\begin{barticle}[mr]
\bauthor{\bsnm{Huang},~\bfnm{Jianhua~Z.}\binits{J.~Z.}},
  \bauthor{\bsnm{Wu},~\bfnm{Colin~O.}\binits{C.~O.}} \AND
  \bauthor{\bsnm{Zhou},~\bfnm{Lan}\binits{L.}}
(\byear{2002}).
\btitle{Varying-coefficient models and basis function approximations for the
  analysis of repeated measurements}.
\bjournal{Biometrika}
\bvolume{89}
\bpages{111--128}.
\bid{doi={10.1093/biomet/89.1.111}, issn={0006-3444}, mr={1888349}}
\end{barticle}
\endbibitem

\bibitem[\protect\citeauthoryear{Larson}{2003}]{larson}
\begin{barticle}[auto:STB|2011-03-03|12:04:44]
\bauthor{\bsnm{Larson},~\bfnm{T.}\binits{T.}}
(\byear{2003}).
\btitle{Why there will be no chain supermarkets in poor inner-city
  neighborhoods}.
\bjournal{California Politics and Policy}
\bvolume{7}
\bpages{22--45}.
\end{barticle}
\endbibitem

\bibitem[\protect\citeauthoryear{Lehmann}{1997}]{Lehmann}
\begin{bbook}[mr]
\bauthor{\bsnm{Lehmann},~\bfnm{E.~L.}\binits{E.~L.}}
(\byear{1997}).
\btitle{Testing Statistical Hypotheses},
\bedition{2nd} ed.
\bpublisher{Springer}, \baddress{New York}.
\bid{mr={1481711}}
\end{bbook}
\endbibitem

\bibitem[\protect\citeauthoryear{Li and Ruppert}{2008}]{Asymptotics}
\begin{barticle}[mr]
\bauthor{\bsnm{Li},~\bfnm{Yingxing}\binits{Y.}} \AND
  \bauthor{\bsnm{Ruppert},~\bfnm{David}\binits{D.}}
(\byear{2008}).
\btitle{On the asymptotics of penalized splines}.
\bjournal{Biometrika}
\bvolume{95}
\bpages{415--436}.
\bid{doi={10.1093/biomet/asn010}, issn={0006-3444}, mr={2521591}}
\end{barticle}
\endbibitem

\bibitem[\protect\citeauthoryear{Liang, Wu and Carroll}{2003}]{Liangetal}
\begin{barticle}[pbm]
\bauthor{\bsnm{Liang},~\bfnm{Hua}\binits{H.}},
  \bauthor{\bsnm{Wu},~\bfnm{Hulin}\binits{H.}} \AND
  \bauthor{\bsnm{Carroll},~\bfnm{Raymond~J.}\binits{R.~J.}}
(\byear{2003}).
\btitle{The relationship between virologic and immunologic responses in AIDS
  clinical research using mixed-effects varying-coefficient models with
  measurement error}.
\bjournal{Biostatistics}
\bvolume{4}
\bpages{297--312}.
\bid{doi={10.1093/biostatistics/4.2.297}, issn={1465-4644}, pii={4/2/297},
  pmid={12925523}}
\end{barticle}
\endbibitem

\bibitem[\protect\citeauthoryear{Lovett et al.}{2002}]{Lovett}
\begin{barticle}[auto:STB|2011-03-03|12:04:44]
\bauthor{\bsnm{Lovett},~\bfnm{A.}\binits{A.}},
  \bauthor{\bsnm{Haynes},~\bfnm{R.}\binits{R.}},
  \bauthor{\bsnm{Sunnenberg},~\bfnm{G.}\binits{G.}} \AND
  \bauthor{\bsnm{Gale},~\bfnm{S.}\binits{S.}}
(\byear{2002}).
\btitle{Car travel time and accessibility by bus to general practitioner
  services: A study using patient registers and GIS}.
\bjournal{Soc. Sci. Med.}
\bvolume{55}
\bpages{97--111}.
\end{barticle}
\endbibitem

\bibitem[\protect\citeauthoryear{Marsh and Schilling}{1994}]{Marsh94}
\begin{barticle}[auto:STB|2011-03-03|12:04:44]
\bauthor{\bsnm{Marsh},~\bfnm{M.~T.}\binits{M.~T.}} \AND
  \bauthor{\bsnm{Schilling},~\bfnm{D.~A.}\binits{D.~A.}}
(\byear{1994}).
\btitle{Equity measurement in facility location analysis---a review and
  framework}.
\bjournal{European J. Oper. Res.}
\bvolume{74}
\bpages{1--17}.
\end{barticle}
\endbibitem

\bibitem[\protect\citeauthoryear{Morris and Carroll}{2006}]{MorrisandCarroll}
\begin{barticle}[mr]
\bauthor{\bsnm{Morris},~\bfnm{Jeffrey~S.}\binits{J.~S.}} \AND
  \bauthor{\bsnm{Carroll},~\bfnm{Raymond~J.}\binits{R.~J.}}
(\byear{2006}).
\btitle{Wavelet-based functional mixed models}.
\bjournal{J. R. Stat. Soc. Ser. B Stat. Methodol.}
\bvolume{68}
\bpages{179--199}.
\bid{doi={10.1111/j.1467-9868.2006.00539.x}, issn={1369-7412}, mr={2188981}}
\end{barticle}
\endbibitem

\bibitem[\protect\citeauthoryear{Morris et al.}{2003}]{Morrisetal}
\begin{barticle}[mr]
\bauthor{\bsnm{Morris},~\bfnm{Jeffrey~S.}\binits{J.~S.}},
  \bauthor{\bsnm{Vannucci},~\bfnm{Marina}\binits{M.}},
  \bauthor{\bsnm{Brown},~\bfnm{Philip~J.}\binits{P.~J.}} \AND
  \bauthor{\bsnm{Carroll},~\bfnm{Raymond~J.}\binits{R.~J.}}
(\byear{2003}).
\btitle{Wavelet-based nonparametric modeling of hierarchical functions in colon
  carcinogenesis}.
\bjournal{J.~Amer. Statist. Assoc.}
\bvolume{98}
\bpages{573--597}.
\bid{doi={10.1198/016214503000000422}, issn={0162-1459}, mr={2011673}}
\bptnote{check related}%
\end{barticle}
\endbibitem

\bibitem[\protect\citeauthoryear{Nychka and Saltzman}{1998}]{space-filling}
\begin{bmisc}[auto:STB|2011-03-03|12:04:44]
\bauthor{\bsnm{Nychka},~\bfnm{D.}\binits{D.}} \AND
  \bauthor{\bsnm{Saltzman},~\bfnm{N.}\binits{N.}}
(\byear{1998}).
\bhowpublished{Design of air quality monitoring networks.
\textit{Lecture Notes in Statist.}
\textbf{132}
51--76.
Springer,
Berlin.}
\end{bmisc}
\endbibitem

\bibitem[\protect\citeauthoryear{Powell et al.}{2007}]{neighfoodstore2}
\begin{barticle}[auto:STB|2011-03-03|12:04:44]
\bauthor{\bsnm{Powell},~\bfnm{L.~M.}\binits{L.~M.}},
  \bauthor{\bsnm{Slater},~\bfnm{S.}\binits{S.}},
  \bauthor{\bsnm{Mirtcheva},~\bfnm{D.}\binits{D.}},
  \bauthor{\bsnm{Bao},~\bfnm{Y.~J.}\binits{Y.~J.}} \AND
  \bauthor{\bsnm{Chaloupka},~\bfnm{F.~J.}\binits{F.~J.}}
(\byear{2007}).
\btitle{Food store availability and neighborhood characteristics in the United
  States}.
\bjournal{Preventive Medicine}
\bvolume{44}
\bpages{189--195}.
\end{barticle}
\endbibitem

\bibitem[\protect\citeauthoryear{Rice and Wu}{2001}]{RiceandWu}
\begin{barticle}[mr]
\bauthor{\bsnm{Rice},~\bfnm{John~A.}\binits{J.~A.}} \AND
  \bauthor{\bsnm{Wu},~\bfnm{Colin~O.}\binits{C.~O.}}
(\byear{2001}).
\btitle{Nonparametric mixed effects models for unequally sampled noisy curves}.
\bjournal{Biometrics}
\bvolume{57}
\bpages{253--259}.
\bid{doi={10.1111/j.0006-341X.2001.00253.x}, issn={0006-341X}, mr={1833314}}
\end{barticle}
\endbibitem

\bibitem[\protect\citeauthoryear{Ruppert}{2002}]{Ruppert}
\begin{barticle}[mr]
\bauthor{\bsnm{Ruppert},~\bfnm{David}\binits{D.}}
(\byear{2002}).
\btitle{Selecting the number of knots for penalized splines}.
\bjournal{J. Comput. Graph. Statist.}
\bvolume{11}
\bpages{735--757}.
\bid{doi={10.1198/106186002321018768}, issn={1061-8600}, mr={1944261}}
\end{barticle}
\endbibitem

\bibitem[\protect\citeauthoryear{Ruppert, Wand and Carroll}{2003}]{semipar}
\begin{bbook}[mr]
\bauthor{\bsnm{Ruppert},~\bfnm{David}\binits{D.}},
  \bauthor{\bsnm{Wand},~\bfnm{M.~P.}\binits{M.~P.}} \AND
  \bauthor{\bsnm{Carroll},~\bfnm{R.~J.}\binits{R.~J.}}
(\byear{2003}).
\btitle{Semiparametric Regression}.
\bseries{Cambridge Series in Statistical and Probabilistic Mathematics}
\bvolume{12}.
\bpublisher{Cambridge Univ. Press}, \baddress{Cambridge}.
\bid{doi={10.1017/CBO9780511755453}, mr={1998720}}
\end{bbook}
\endbibitem

\bibitem[\protect\citeauthoryear{Serban}{2011}]{suppl}
\begin{bmisc}[auto:STB|2011-03-03|12:04:44]
\bauthor{\bsnm{Serban},~\bfnm{N.}\binits{N.}}
(\byear{2011}).
\bhowpublished{Supplement to ``A space--time varying coefficient model: The
  equity of service accessibility.''
  \href{http://dx.doi.org/10.1214/11-AOAS473SUPP}{DOI:10.1214/11-AOAS473SUPP}}.
\end{bmisc}
\endbibitem

\bibitem[\protect\citeauthoryear{Sim and Reid}{1999}]{SimandReid}
\begin{barticle}[pbm]
\bauthor{\bsnm{Sim},~\bfnm{J.}\binits{J.}} \AND
  \bauthor{\bsnm{Reid},~\bfnm{N.}\binits{N.}}
(\byear{1999}).
\btitle{Statistical inference by confidence intervals: Issues of interpretation
  and utilization}.
\bjournal{Phys. Ther.}
\bvolume{79}
\bpages{186--195}.
\bid{issn={0031-9023}, pmid={10029058}}
\end{barticle}
\endbibitem

\bibitem[\protect\citeauthoryear{Small and McDermott}{2006}]{princeton}
\begin{bmisc}[auto:STB|2011-03-03|12:04:44]
\bauthor{\bsnm{Small},~\bfnm{M.~L.}\binits{M.~L.}} \AND
  \bauthor{\bsnm{McDermott},~\bfnm{M.}\binits{M.}}
(\byear{2006}).
\bhowpublished{The presence of organizational resources in poor urban
  neighborhoods: An analysis of average and contextual effects. \textit{Social
  Forces} \textbf{84} 1697}.
\end{bmisc}
\endbibitem

\bibitem[\protect\citeauthoryear{Staicu, Crainiceanu and Carroll}{2010}]{Staicu}
\begin{barticle}[pbm]
\bauthor{\bsnm{Staicu},~\bfnm{Ana-Maria}\binits{A.-M.}},
  \bauthor{\bsnm{Crainiceanu},~\bfnm{Ciprian~M.}\binits{C.~M.}} \AND
  \bauthor{\bsnm{Carroll},~\bfnm{Raymond~J.}\binits{R.~J.}}
(\byear{2010}).
\btitle{Fast methods for spatially correlated multilevel functional data}.
\bjournal{Biostatistics}
\bvolume{11}
\bpages{177--194}.
\bid{doi={10.1093/biostatistics/kxp058}, issn={1468-4357}, pii={kxp058},
  pmcid={2830578}, pmid={20089508}}
\end{barticle}
\endbibitem

\bibitem[\protect\citeauthoryear{Talen}{1997}]{Talen}
\begin{barticle}[auto:STB|2011-03-03|12:04:44]
\bauthor{\bsnm{Talen},~\bfnm{E.}\binits{E.}}
(\byear{1997}).
\btitle{The social equity of urban service distribution: An exploration of park
  access in Pueblo, Co and Macon, GA}.
\bjournal{Urban Geography}
\bvolume{18}
\bpages{521--541}.
\end{barticle}
\endbibitem

\bibitem[\protect\citeauthoryear{Talen}{2001}]{accessschool}
\begin{barticle}[auto:STB|2011-03-03|12:04:44]
\bauthor{\bsnm{Talen},~\bfnm{E.}\binits{E.}}
(\byear{2001}).
\btitle{School, community, and spatial equity: An empirical investigation of
  access to elementary schools in West Virginia}.
\bjournal{Annals of the Association of American Geographers}
\bvolume{91}
\bpages{465--486}.
\end{barticle}
\endbibitem

\bibitem[\protect\citeauthoryear{Talen and Anselin}{1998}]{assessacess}
\begin{bmisc}[auto:STB|2011-03-03|12:04:44]
\bauthor{\bsnm{Talen},~\bfnm{E.}\binits{E.}} \AND
  \bauthor{\bsnm{Anselin},~\bfnm{L.}\binits{L.}}
(\byear{1998}).
\bhowpublished{Assessing spatial equity: An evaluation of measures of
  accessibility to public playgrounds. \textit{Environment and Planning A}
  \textbf{30} 595--613}.
\end{bmisc}
\endbibitem

\bibitem[\protect\citeauthoryear{Wahba}{1990}]{Wahba}
\begin{bbook}[mr]
\bauthor{\bsnm{Wahba},~\bfnm{Grace}\binits{G.}}
(\byear{1990}).
\btitle{Spline Models for Observational Data}.
\bseries{CBMS-NSF Regional Conference Series in Applied Mathematics}
\bvolume{59}.
\bpublisher{SIAM},
  \baddress{Philadelphia, PA}.
\bid{mr={1045442}}
\end{bbook}
\endbibitem

\bibitem[\protect\citeauthoryear{Waller et al.}{2007}]{waller07}
\begin{barticle}[mr]
\bauthor{\bsnm{Waller},~\bfnm{Lance~A.}\binits{L.~A.}},
  \bauthor{\bsnm{Zhu},~\bfnm{Li}\binits{L.}},
  \bauthor{\bsnm{Gotway},~\bfnm{Carol~A.}\binits{C.~A.}},
  \bauthor{\bsnm{Gorman},~\bfnm{Dennis~M.}\binits{D.~M.}} \AND
  \bauthor{\bsnm{Gruenewald},~\bfnm{Paul~J.}\binits{P.~J.}}
(\byear{2007}).
\btitle{Quantifying geographic variations in associations between alcohol
  distribution and violence: A comparison of geographically weighted regression
  and spatially varying coefficient models}.
\bjournal{Stoch. Environ. Res. Risk Assess.}
\bvolume{21}
\bpages{573--588}.
\bid{doi={10.1007/s00477-007-0139-9}, issn={1436-3240}, mr={2380676}}
\end{barticle}
\endbibitem

\bibitem[\protect\citeauthoryear{Wood}{2006}]{wood}
\begin{bbook}[mr]
\bauthor{\bsnm{Wood},~\bfnm{Simon~N.}\binits{S.~N.}}
(\byear{2006}).
\btitle{Generalized Additive Models: An Introduction with R}.
\bpublisher{Chapman \& Hall/CRC}, \baddress{Boca Raton, FL}.
\bid{mr={2206355}}
\end{bbook}
\endbibitem

\bibitem[\protect\citeauthoryear{Wu and Liang}{2004}]{WuandLiang}
\begin{barticle}[auto:STB|2011-03-03|12:04:44]
\bauthor{\bsnm{Wu},~\bfnm{H.}\binits{H.}} \AND
  \bauthor{\bsnm{Liang},~\bfnm{H.}\binits{H.}}
(\byear{2004}).
\btitle{Backfitting random varying-coefficient models with timedependent
  smoothing covariates}.
\bjournal{Scand. J. Statist.}
\bvolume{31}
\bpages{3--20}.
\end{barticle}
\endbibitem

\bibitem[\protect\citeauthoryear{Wu and Zhang}{2002}]{WuandZhang}
\begin{barticle}[mr]
\bauthor{\bsnm{Wu},~\bfnm{Hulin}\binits{H.}} \AND
  \bauthor{\bsnm{Zhang},~\bfnm{Jin-Ting}\binits{J.-T.}}
(\byear{2002}).
\btitle{Local polynomial mixed-effects models for longitudinal data}.
\bjournal{J. Amer. Statist. Assoc.}
\bvolume{97}
\bpages{883--897}.
\bid{doi={10.1198/016214502388618672}, issn={0162-1459}, mr={1941417}}
\end{barticle}
\endbibitem

\bibitem[\protect\citeauthoryear{Zenk et al.}{2005}]{zenk}
\begin{barticle}[pbm]
\bauthor{\bsnm{Zenk},~\bfnm{Shannon~N.}\binits{S.~N.}},
  \bauthor{\bsnm{Schulz},~\bfnm{Amy~J.}\binits{A.~J.}},
  \bauthor{\bsnm{Israel},~\bfnm{Barbara~A.}\binits{B.~A.}},
  \bauthor{\bsnm{James},~\bfnm{Sherman~A.}\binits{S.~A.}},
  \bauthor{\bsnm{Bao},~\bfnm{Shuming}\binits{S.}} \AND
  \bauthor{\bsnm{Wilson},~\bfnm{Mark~L.}\binits{M.~L.}}
(\byear{2005}).
\btitle{Neighborhood racial composition, neighborhood poverty, and the spatial
  accessibility of supermarkets in metropolitan Detroit}.
\bjournal{Am. J. Public Health}
\bvolume{95}
\bpages{660--667}.
\bid{doi={10.2105/AJPH.2004.042150}, issn={0090-0036}, pii={95/4/660},
  pmcid={1449238}, pmid={15798127}}
\end{barticle}
\endbibitem

\bibitem[\protect\citeauthoryear{Zhang}{2004}]{varying7}
\begin{barticle}[mr]
\bauthor{\bsnm{Zhang},~\bfnm{Daowen}\binits{D.}}
(\byear{2004}).
\btitle{Generalized linear mixed models with varying coefficients for
  longitudinal data}.
\bjournal{Biometrics}
\bvolume{60}
\bpages{8--15}.
\bid{doi={10.1111/j.0006-341X.2004.00165.x}, issn={0006-341X}, mr={2043613}}
\end{barticle}
\endbibitem


\end{thebibliography}
\end{document}